\documentclass[12pt, draftclsnofoot, onecolumn]{IEEEtran}

\def\ignore#1{}

\usepackage{amsmath,epsfig,epsf,psfrag,amssymb,amsfonts,latexsym,amsmath,color}

\usepackage{graphicx}
\usepackage{caption}
\usepackage{epstopdf}
\DeclareGraphicsExtensions{.eps}

\usepackage{bbm}
\usepackage{verbatim}
\usepackage{cite,enumerate}
\usepackage{srcltx,cases}
\usepackage[mathscr]{eucal}
\usepackage{url}
\usepackage{hyperref}

\usepackage{subcaption}

\newtheorem{theorem}{Theorem}
\newtheorem{lemma}{Lemma}

\newtheorem{corollary}{Corollary}
\newtheorem{remark}{\textit{Remark}}

\newtheorem{example}{Example}

\newcommand{\beq}{\begin{equation}}
\newcommand{\eeq}{\end{equation}}

\begin{document}

\title{Scaling Laws for Ergodic Spectral Efficiency in MIMO Poisson Networks }

\author{  \IEEEauthorblockN{Junse Lee, Namyoon Lee and Fran\c{c}ois Baccelli
	 } 
	\thanks{J. Lee and F. Baccelli are with the Wireless Networking and Communications Group, Department of Electrical and Computer Engineering, The University of Texas at
Austin, Austin, TX 78712 USA (e-mali: junselee@utexas.edu and francois.baccelli@austin.utexas.edu)
	N. Lee  is with the Department of Electrical Engineering, POSTECH, Pohang, Gyeongbuk, Korea 37673 (e-mail:nylee@postech.ac.kr). }}

\date{}
\maketitle

\begin{abstract}
%	In this paper, we consider a large wireless random network and investigate the gain of multiple transmit and receive antennas. The transmit-and-receive node pairs communicate in a certain range and under a common shared spectrum. The transmit nodes are distributed according to a homogeneous Poisson point process, allowing expressions for the system performance metrics. The channel state information is only available at each receiver and some simple linear receiver architectures are used to nullifying the inter-stream and inter-node interferences. One key finding is that the multiplexing gain of the sum spectral efficiency per unit area is obtained under a certain region of transmit and receive antenna numbers and system model parameters even the nodes are very densely distributed. 	

In this paper, we examine the benefits of multiple antenna communication in random wireless networks, the topology of which is modeled by stochastic geometry. The setting is that of the Poisson bipolar model introduced in \cite{baccelli2009stochasticvol2}, which is a natural model for ad-hoc and device-to-device (D2D) networks.  The primary finding is that, with knowledge of channel state information between a receiver and its associated transmitter, by zero-forcing successive interference cancellation, and for appropriate antenna configurations, the ergodic spectral efficiency can be made to scale linearly with both 1) the minimum of the number of transmit and receive antennas, 2) the density of nodes and 3) the path-loss exponent. 
This linear gain is achieved by using the transmit antennas to send multiple data streams (e.g. through an open-loop transmission method) and by exploiting the receive antennas to cancel interference. Furthermore, when a receiver is able to learn channel state information from a certain number of near interferers, higher scaling gains can be achieved when using a successive interference cancellation method. A major implication of the derived scaling laws is that spatial multiplexing transmission methods are essential for obtaining better and eventually optimal scaling laws in multiple antenna random wireless networks. Simulation results support this analysis.

	% $need$ $to$ $add$ $more$ $when$ $we$ $find$ $more$ $results.$
\end{abstract}

\IEEEpeerreviewmaketitle
%%%%%%%%%%%%%%%%%%%%%%%%%%%%%%%%%%%%%%%%%%%%%%%%%%%%%%%%%%%%%%%%%%%%%%%%%%%%%%%%%%%%%%%%%%%
\section{Introduction}\label{sec:intro}

A multiple-input-multiple-output mobile ad hoc network (MIMO-MANET) is an infrastructure-less network
in which a large number of transmit-and-receive pairs, each with multiple antennas,
communicate by sharing some common spectrum \cite{gupta2000capacity,toumpis2003capacity}.   
Such networks are fundamental in a variety of applications including
car-to-car and device-to-device communication systems
\cite{hartenstein2008tutorial,doppler2009device,fodor2012design}.
It is therefore of great importance to characterize the system-level performance of 
such networks\cite{blum2003mimo,chen2006mimo,ozgur2007hierarchical}. 

Despite extensive research over a few decades, analytical expressions for the spectral efficiency
of such systems are still missing. The principal difficulty has been the lack of a tractable
model quantifying uncoordinated inter-node interference together with inter-stream interference at
a receiver equipped with multiple antennas. In this paper, we leverage two analytical tools to
cope with this difficulty. The first one is stochastic geometry which models the 
locations of links as Poisson dipoles \cite{baccelli2009stochasticvol2} and allows one
to compute the distribution of the interference power. 
The second one is random matrix theory \cite{tulino2004random}, which is exploited 
for calculating the distribution of inter-stream interference power under different MIMO detection techniques.
Combining these tools, we characterize the ergodic spectral efficiencies and the scaling laws
of a super-dense MIMO-MANET system, under Poisson assumptions on the node locations,
and when considering two major types of channel knowledge at receivers.
By leveraging the closed-form expressions which are derived, we highlight the
interplay among four key system parameters determining the scaling laws,
namely the number of antennas at the transmitter,
the number of antennas at the receiver, the node density, and the path-loss exponent. 

\subsection{Related Works}

There has been extensive work on the capacity of MIMO-MANETs. MIMO-MANETs can be modeled
as MIMO interference networks in which a finite number of transmit-and-receiver
pairs communicate by sharing the same spectrum, without transmitter cooperation.
\cite{blum2003mimo} studied the capacity of a MIMO-MANET by treating inter-node
interference as additional noise at a receiver,
and derived the optimal power allocation strategy for the MIMO transmission.
For instance, in a certain range of interference-to-noise ratios,
it turns out that allocating the whole power to one antenna
(i.e., using a single stream transmission) is optimal.
\cite{chen2006mimo} and \cite{yu2005capacity} extended the result of \cite{blum2003mimo},
and demonstrated that the asymptotic spectral efficiency is improved by
sending multiple data streams. A common assumption of these studies is that
the distances between any two nodes in the network are deterministic
\cite{blum2003mimo} or identical \cite{chen2006mimo},
which is unrealistic to model MANETs in practice.
This approach cannot be used to assess which MIMO
transmission techniques provide the highest gains in large random MANETs.

When considering more realistic random network topology assumptions,
the rates achievable in MANETs have been studied in
\cite{gupta2000capacity,franceschetti2007closing,leveque2004information,grossglauser2001mobility,negi2004capacity,ozgur2007hierarchical,franceschetti2009capacity}. 
The study of scaling laws within this context was initiated
by Gupta and Kumar's seminal paper \cite{gupta2000capacity}.
Under the assumption that $n$ nodes are randomly located in the unit disk,
Gupta and Kumar showed that multihop routing based on a decode-and-forward 
scheme can reach to a total throughput which scales as $\mathcal{O}(\sqrt{n})$.
%In \cite{kulkarni2004deterministic}, the same results with a general analysis technique were stated. 
By using percolation theory, it was later shown in \cite{franceschetti2007closing}
that a better scaling law of order $\mathcal{O}(\sqrt{n/\log{n}})$ is achievable.
Subsequently, improved scaling results were derived in MANETs, assuming that
some specific additional assumptions hold on mobility \cite{grossglauser2001mobility}, bandwidth \cite{negi2004capacity}, or node-cooperation \cite{ozgur2007hierarchical}.
The main differences between our work and this line of research are 
the following: 
(1) our model is based on Poisson dipoles and assumes that 
source-destination pairs communicate with each other relying
upon single-hop transmissions, i.e., neither multi-hop routing schemes
nor node-cooperation are allowed (in a sense, the present paper is more
focussed on D2D than on MANETs).
(2) we focus on the use of multiple antennas at both transmitters 
and receivers, while this line
of research was centered on the scenario
with a single antenna at both transmitters and receivers.
(3) our performance metric is spatially-averaged ergodic spectral
efficiency, while the work alluded to above focused
on transport capacity. (4) even if new scaling laws are our main results,
our approach also provides exact formulas for the mean Shannon rate of a typical link
and the spectral efficiency per unit area
(see e.g.Theorems b\ref{theo:Direct_ZF} and \ref{theo:Direct_ZFSIC} below), and
goes hence beyond the scaling law setting.
%, packets route with a multi-hop fashion from their sources to destinations and source-destination pairs are randomly chosen in random networks. Also, a performance metric called transport capacity which is defined by the aggregated bit meters per second over network is considered. However, the viewpoint of this paper is somewhat different with these papers. We are more focused on the rate analysis and its scaling laws under one-hop large dense wireless network with fixed source-destination pairs.

In the present paper, we assume that the interferer
locations are Poisson distributed over the plane
\cite{baccelli2009stochastic,Stoyan}, which is an appropriate model
for e.g. D2D, where transmitters are randomly located in
an uncoordinated manner. Using this model, the transmission capacity
of ad hoc networks, which quantifies the maximum allowable
spatial density of successful transmissions per unit area,
subject to a given outage probability constraint,
was characterized in certain settings. For example, the
transmission capacity expressions of ad hoc networks
were found when adopting spread spectrum techniques 
\cite{weber2005transmission,andrews2007ad},
interference cancellation  \cite{weber2007transmission,blomer2009transmission,zhang2014performance}, 
and multiple-antenna transmission methods \cite{hunter2008transmission,jindal2011multi,akoum2011spatial,louie2011open,huang2012spatial,vaze2012transmission,kountouris2009transmission,lee2015spectral}.
In particular, in \cite{jindal2011multi}, it was demonstrated that
interference cancellation techniques at a receiver employing 
multiple antennas can provide a linear increase of the
transmission capacity of ad hoc networks with the node density.
In \cite{vaze2012transmission}, it was shown that for a MIMO setting,
a single stream transmission is optimal in terms of 
transmission capacity, when all the degrees of freedom of
the receive antennas are used for interference cancellation. %This is doubt on the use of spatial multiplexing transmissions in the MIMO ad hoc networks.

Arguably, a common shortcoming of the transmission capacity metric is
that it cannot capture the effects of rate adaptation techniques,
which are the key features used in many modern wireless systems 
to track and exploit channel variations \cite{lozano2012yesterday}.
The main novelty of the present paper compared to this line of thought
is the analysis of the ergodic spectral efficiency (rather than transport capacity),
which quantifies the achievable Shannon transmission rate per unit
area when adapting the rate to the different local conditions.
For a single-input-multiple-output (SIMO) setting,
the recent work in \cite{lee2016spectral} showed that the sum
spectral efficiency per link can increase linearly with
both the density and the path loss exponent provided
the number of antennas is a linear function of the density.
For a MIMO setting, however, it is still unknown whether
\emph{spatial multiplexing} transmission techniques
\cite{tse2005fundamentals} can improve the scaling laws of
the sum spectral efficiency. We recall that
\emph{spatial multiplexing} consists in transmitting
different data streams on the transmit antennas and
in identifying/discriminating between these streams at the receiver,
while \emph{transmit diversity} consists in sending
the same data symbols over multiple transmit antennas
to enhance the reliability. The main qualitative achievement of this paper
is a proof that the answer to this question is positive
and more precisely the identification of the
network densities and antenna configurations for which
\emph{spatial multiplexing} strategies achieve higher
sum spectral efficiency per unit area than the methods
based on \emph{transmit diversity}.

\subsection{Main Contributions}

We consider a random network the topology of
which modeled by %homogeneous Poisson point process (PPP) $\rightarrow$ 
a \emph{Poisson bipolar network} \cite{baccelli2009stochasticvol2}
with density $\lambda$ on $\mathbb{R}^2$.
In this model, each transmitter has its receiver at some random distance.
Each transmitter is equipped with $N_{\rm t}$ antennas and is
assumed to send $N_{\rm t}$ data streams to its associated receiver, 
equipped with $N_{\rm r} (\geq N_{\rm t})$ antennas.
Our key findings can be summarized as follows:

\begin{itemize}
	\item We first consider the case where each receiver has knowledge
of the state of the channel between its transmitter and itself only.
We refer to this channel knowledge assumption as {\em direct channel state information} 
(DCSI) at receiver (DCSIR). Under the premise of this channel knowledge,
and under zero-forcing (ZF) detection and ZF-based successive interference cancellation
(ZF-SIC) detection respectively,
we derive analytical expressions of the sum spectral efficiency as a function of
1) the network density $\lambda$, 2) the number of transmit and receive antennas
($N_{\rm t}$ and $N_{\rm r}$), 3) the path-loss exponent $\alpha$, and 4)
the signal-to-noise ratio (SNR). By deriving a closed form lower and upper bound
on this sum spectral efficiency, we show that, as $\lambda$ goes to infinity,
when $N_{\rm t}=c_1\lambda^{\beta_1}$, $N_{\rm r}=c_2\lambda^{\beta_2}$
for some constants $c_1,c_2>0$, $\beta_1\leq\beta_2$ and $\alpha>2$,
the scaling laws of the ergodic spectral efficiency per link is
	\begin{align}
	\Theta(\lambda^{\beta_1}\log_2(1+\lambda^{\beta_2-\beta_1-\frac{\alpha}{2}})),
	\end{align} 
for both ZF and ZF-SIC. One important implication of this scaling law is that 
when $\beta_2 \geq \beta_1 + \frac{\alpha}{2}$, transmitting multiple streams is
more beneficial in the scaling law sense than sending a single stream,
which strongly contrasts with the result derived on transmission capacity \cite{vaze2012transmission}.
In fact, this result agrees with the intuition that it should be possible
to improve the data rates per link by having $N_{\rm t}=c_1\lambda^{\beta_1}$
and by transmitting multiple data streams (\emph{multiplexing gain}),
provided the remaining degrees of freedom at the receiver are sufficient
to cancel both inter-stream interference and inter-node interference and 
to discriminate between the independent data streams.
Furthermore, this scaling law expression generalizes
the result for the SIMO case derived in \cite{lee2016spectral} to the MIMO case.

\item We also consider the case where each receiver is able to learn the CSI
of its $L$-nearest interferers with
($0<L\leq\lfloor\frac{N_{\rm r}}{N_{\rm t}}\rfloor-1$)%
\footnote{$\lfloor x \rfloor$ denotes the largest integer no more than $x$.},
which is referred to here as local CSIR (LCSIR).
Using a ZF-SIC detection technique for suppressing both inter-stream and inter-node
interference, we give an exact expression of the sum spectral efficiency.
By leveraging this expression, we get an achievable scaling law of the sum
spectral efficiency per link of the form: 
	\begin{align}
	\Omega(\lambda^{\beta_1}(\log_2(1+\lambda^{(\beta_2-\beta_1-1)\frac{\alpha}{2}}))),
	\end{align}
when $N_{\rm t}=c_1\lambda^{\beta_1}$ and $N_{\rm r}=c_2\lambda^{\beta_2}$,
for some constants $c_1,c_2>0$ and $\beta_1\leq\beta_2$, $\alpha>2$, and for
$L=\lfloor\frac{N_{\rm r}}{N_{\rm t}}\rfloor-1$. This result also demonstrates that the
MIMO transmission method improves the scaling law of the ergodic spectral efficiency
per link by increasing multiplexing gains, provided $\beta_2\geq \beta_1+1$.
Comparing to DCSIR, with LCSIR, it is possible to increase the sum spectral
efficiency with both the path-loss exponent and the number of transmit antennas.
This multiplicative gain in the achievable scaling law comes
from the fact that the receiver exploits LCSIR.

	%\item In contrast to the previous researches, we can expect the multiplexing gain in the scaling law by increasing the number of transmit data streams. This is not obtained when the performance metric is transmission capacity.
%	Furthermore, for given $\beta_2$, and $\alpha$, we obtain the optimal $\beta_1$ in terms of the scaling law sense. This result gives the information-theoretical guide for the optimal number of transmit data streams. The optimal $\beta_1$ is $\beta_2-\frac{\alpha}{2}$ and the corresponding scaling law is $\Theta(\lambda^{\beta_2-\frac{\alpha}{2}})$ when the direct channel information is only available at each receiver, and $\beta_2-1$ and the corresponding scaling law is 
\end{itemize}

This paper is organized as follows. The network model, the performance metrics,
and the receiver schemes are discussed in Section \ref{sec:system model}.
The exact expression and the scaling law for ergodic spectral efficiency
are provided in Section \ref{sec:Direct_CSIR} in the DCSIR case and
in Section \ref{sec:local_CSIR} in the LCSIR case. We conclude in Section \ref{sec:conclu}.

%%%%%%%%%%%%%%%%%%%%%%%%%%%%%%%%%%%%%%%%%%%%%%%%%%%%%%%%%%%%%%%%%%%%%%%%%%%%%%%%%%%%%%%%%%%
\section{System Model}\label{sec:system model}

\subsection{Network Model}
\begin{figure}
	\begin{center}
		\includegraphics[width = 4in,height=3.7in]{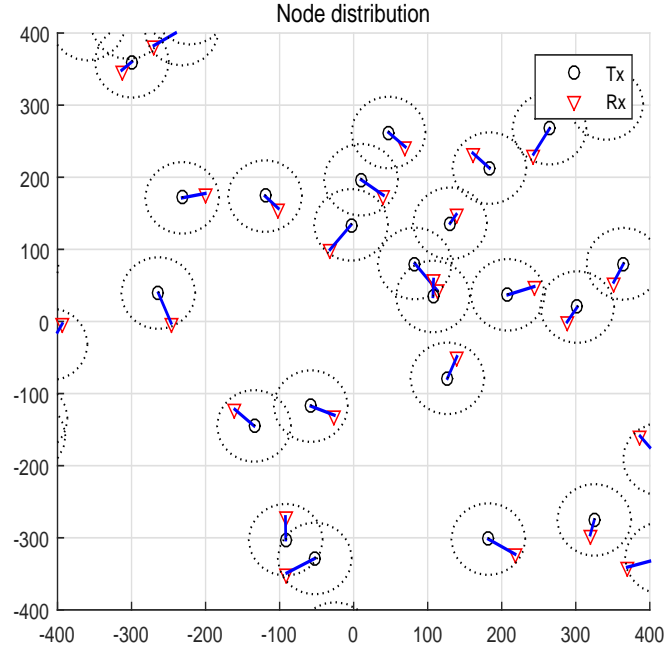}%\epsfxsize=4in {\epsfbox{Network_model.eps}}
		\caption{\small A snapshot of bipolar MANET where $\lambda=0.00004/m^2$ and $R_{ d} = 50m$. }
		\label{fig:network model}
	\end{center}
\end{figure}

We consider a \emph{Poisson bipolar network model} which features an infinite 
number of transmitter-receiver pairs scattered in the Euclidean plane.
Let $\Phi=\{X_{ i}\}_{i\in\mathbb{N}}$ denote the locations of the
transmitters, which are assumed to form some realization of a
homogeneous PPP with positive and finite intensity $\lambda$ on $\mathbb{R}^2$.
Let $\bar{\Phi}=\{Y_{ i}\}_{i\in\mathbb{N}}$ denote the locations of
the receivers. The receiver $Y_{ i}$ of $X_i$ is assumed to
be uniformly distributed on a ring with inner radius 1
and outer radius $R_{ d}$ centered at $\{X_{ i}\}$, where $R_d>1$. 
Fig.~\ref{fig:network model} provides a snapshot of network topology
with $R_{ d}=50m$ and $\lambda = 0.000004/m^2$.
We assume that each receiver is equipped with $N_{\rm r}$ antennas,
whereas transmitters have a random number $N_{\rm t}$
in $[1,N_{\rm r}]\in\mathbb{N}$ of antennas.
We denote the probability of having $k$ transmit antennas
by $p_{ k}$ where $\sum_{{ k}=1}^{N_{\rm r}}p_{ k}=1$. 
These numbers of transmit are assumed independent and
identically distributed (i.i.d.) over links.

\subsection{Signal Model}

A transmitter $X_{ k}\in\Phi$ communicates with its associated receiver $Y_{ k}$,
and sends a signal $\mathbf{s}_{ k}\in\mathbb{C}^{N_{{\rm t},k}\times 1}$
when $X_{ k}$ has $N_{{\rm t},k}$ antennas, with power constraint
$\mathbb{E}[\|\mathbf{s}_k\|^2]= P$. Here, we assume the transmit power is 
equally allocated to all antennas. Assuming a frequency-flat channel,
the received signal at the ${ k}$-th receiver, $\mathbf{y}_{ k}\in\mathbb{C}^{N_{\rm r}\times 1}$ is
\begin{equation}\label{eq:received_vector}
\mathbf{y}_{ k}=\sum_{l,X_l\in\Phi}d_{ k,l}^{-\frac{\alpha}{2}}\mathbf{H}_{ k,l}\mathbf{s}_{ l}+\mathbf{z}_{ k}\mbox{,}
\end{equation}
where $\mathbf{H}_{ k,l}\in\mathbb{C}^{N_{\rm r}\times N_{{\rm t},l}}$
is the channel matrix and $d_{ k,l}$ the distance from $X_l$ to $Y_{ k}$, respectively.
Moreover, $\mathbf{z}_k\in\mathbb{C}^{N_{\rm r}\times 1}$ is
the noise vector at receiver $Y_{ k}$. Furthermore, we assume that all entries of
$\mathbf{H}_{ k,l}$ are i.i.d. complex Gaussian random variables with zero mean
and unit variance, i.e.\, $\mathcal{CN}(0,1)$, and that all entries of
$\mathbf{z}_{ k}$ are i.i.d.\ $\mathcal{CN}(0,\sigma^2)$, where $\sigma^2$ is the noise variance.

\subsection{Receive Filters and Performance Metrics}
We assume that receiver $Y_{ k}$ can measure CSI from its
associated transmitter $X_{ k}$ and from the $L_{ k}$ nearest transmitters, i.e.\,
$\{X_{ k_i}\}_{i=1}^{L_{ k}}$, where
$0\leq L_{ k}\leq \max\{n|\sum_{i=1}^{n}N_{{\rm t},j_i}\leq N_{\rm r}-N_{{\rm t},k}\}$%
\footnote{With this condition, the number of received data streams at $Y_k$ is no larger than $N_{\rm r}$.
This assumption is necessary for decoding the independent data streams in ZF and ZF-SIC.
If all transmitters are equipped with $N_{\rm t}$ antennas,
$L_k=\lfloor \frac{N_{\rm r}}{N_{\rm t}} \rfloor -1,\forall k\in\mathcal{K}$.
Further, we denote the $j$-th nearest interferer from $Y_k$ by $X_{k_j}$.}.
 %It is able to construct an augmented channel matrix $\mathbf{H}_k=[\mathbf{H}_{k,k},\mathbf{H}_{k,k_1},\ldots,\mathbf{H}_{k,k_L}]\in\mathbb{C}^{N_{\rm r}\times(1+L)N_{\rm t}}$. 
It will be assumed that $X_{ k}$ sends $N_{{\rm t},k}$ data streams without using any precoding,
i.e., that an open-loop MIMO transmission is used, and also that the receiver
uses linear receive filters to detect the desired data symbol to eliminate the 
\emph{inter-stream interference} and the \emph{inter-node interference}. 

Let $\mathbf{v}_{ k}(m)\in\mathbb{C}^{N_{\rm r}\times 1}$, $m=1,\ldots, N_{{\rm t},k},$
denote the receive filter vector used at $Y_k$ for detecting the $m$-th data stream of its transmitter.
Then, the resulting signal-to-interference-and-noise ratio (SINR) for the $m$-th data stream 
of the $k$-th link is
\begin{align}
\mbox{SINR}_{ k}(m)=\frac{H_{ k,k}(m)d_{ k,k}^{-\alpha}}{I_{ k1}(m)+I_{ k2}(m)+I_{ k3}(m)+\frac{N_{{\rm t},k}\sigma^2}{P}}\mbox{,} \label{eq:SINR}
\end{align}
where
\begin{eqnarray*}
& H_{k,k}(m)=\|\mathbf{v}_{ k}^*(m)\mathbf{H}_{ k,k}(:,m)\|^2, 
& I_{ k1}(m)=\sum_{i=1,\ldots,N_{{\rm t},k}}^{i\neq m}\|\mathbf{v}_{ k}^*(m)\mathbf{H}_{ k,k}(:,i)\|^2d_{ k,k}^{-\alpha}, \\
& I_{ k2}(m)=\sum_{l\in\{{k_1},{k_2},\ldots,{k_{L_{ k}}}\}}\|\mathbf{v}_{ k}^*(m)\mathbf{H}_{ k,l}\|^2d_{ k,l}^{-\alpha}, 
& I_{ k3}(m)=\hspace{-.5cm}\sum_{l,X_l\in\{\Phi\setminus X_k,X_{k_1},X_{k_2},\ldots,X_{k_{L_{ k}}}\}}\|\mathbf{v}_{ k}^*(m)\mathbf{H}_{ k,l}\|^2d_{ k,l}^{-\alpha}.
\end{eqnarray*}
Here, we denote the conjugate transpose by $*$ and the $i$-th column of
the matrix $A$ by $A(:,i)$. As can be seen in (\ref{eq:SINR}),
the total amount of interference at the receiver can be decomposed 
into three factors: 1) the \emph{inter-stream interference} $I_{k1}$, 
2) the \emph{inter-node interference} from the $L_{ k}$-dominant interferers, $I_{k2}$,
and 3) the \emph{inter-node interference}, $I_{k3}$, which is the interference from the other nodes.
Then, the achievable rate of the $k$-th link is 
\begin{align}
C_k =\sum_{m=1}^{N_{{\rm t},k}}C_k(m)=\sum_{m=1}^{N_{{\rm t},k}}
\log_{2}(1+\mbox{SINR}_k(m)).
\end{align}

The main target performance metric in this paper is \emph{ergodic spectral efficiency}. 
%The spectral efficiency is the information rate that can be transmitted over a given bandwidth through the physical channel. Here, the ergodicity in this paper is over the time-domain process, or averaging over the small-scale multipath fading. However, by leveraging the Palm theory \cite{baccelli2009stochastic} under our stationary PPP framework, it is possible to compute the average of the spectral efficiency over both the time and space domain. }
The sum spectral efficiency per unit area is defined by
\begin{align}\label{eq:palm}
C=\frac{1}{|\mathcal{A}|}C_{\mathcal{A}}=\frac{1}{|\mathcal{A}|}\mathbb{E}\left[\sum_{k,X_k\in\mathcal{A}}C_k\right]
=\lambda\mathbb{E}^0\left[\sum_{m=1}^{N_{{\rm t},0}}\log_2(1+\mathrm{SINR}_0(m))\right]\mbox{,}
\end{align}
where for any $\mathcal{A}\subset \mathbb{R}^2$,
$|\mathcal{A}|$ is the area of $\mathcal{A}$, 
$C_{\mathcal{A}}$ is the sum spectral efficiency of $\mathcal{A}$,
and $\mathbb{E}^0$ denotes the Palm expectation
\cite{baccelli2009stochastic} of the receiver PPP.
The fact that the last expression does not
depend on the choice of $\mathcal{A}$ results from the
stationarity assumptions \cite{baccelli2009stochastic}.
Here, $\mbox{SINR}_{0}$ denotes the SINR measured at
the receiver located at the origin%
\footnote{By \emph{Slivnyak's theorem}\cite{baccelli2009stochastic},
it is possible locate the typical receiver at the origin.
We label the typical transmitter and the typical receiver
by $X_0$ and $Y_0=0$, respectively.
The distance between $Y_0$ and $\{X_k\}_{k\in\{0\}\cup\mathbb{N}}$, 
the channel matrix $\mathbf{H}_{0,l},l\in\{0\}\cup\mathbb{N}$,
the linear receiver filter $\mathbf{v}_0(m)$, $I_0(m)$, and the $j$-th nearest interferers from $Y_0$, i.e., $X_{0_j}$ are defined similarly.}.
%and \eqref{eq:palm} holds by the \emph{Campbell-Matthes formula}\cite{baccelli2009stochastic} for a stationary PPP. 
Furthermore, the {\em spectral efficiency of the typical link},
or equivalently the \emph{spectral efficiency} per link is defined by
\begin{align}
C_{\rm link}=\frac{1}{\lambda}C = \mathbb{E}^0\left[\sum_{m=1}^{N_{{\rm t},0}}\log_2(1+\mathrm{SINR}_0(m))\right]\mbox{.}
\end{align}
Here, for the above quantities, we will use the terms
ergodic spectral efficiency or ergodic spectral efficiency per link, respectively. 
The ergodicity is over both the time-domain 
(averaging over the small-scale multipath fading)
and over space (averaging over all Poisson configurations).

%We assume that the probability density function (pdf)
%of $N_{{\rm t},0}$ is $\mathbb{P}[N_{{\rm t},0}=k]=p_k$.
We will denote the sum spectral efficiency per unit area
by $C^{\rm ZF}$ under ZF, and by $C^{\rm SIC}$ undre ZF-SIC,
the sum spectral efficiency of {$\mathcal{A}$} with ZF
by $C_{\mathcal{A}}^{\rm ZF}$ under ZF and
by $C_{\mathcal{A}}^{\rm SIC}$ under ZF-SIC, and
the spectral efficiency per link
by $C_{\rm link}^{\rm ZF}$ under ZF
by $C_{\rm link}^{\rm SIC}$ under ZF-SIC.

\subsubsection{ZF detection} 
The main idea of the \emph{ZF-decorrelator}\cite{tse2005fundamentals}
is to construct $\mathbf{v}_{ k}(m)$ so as to remove both
$I_{k1}(m)$ and $I_{k2}(m)$ simultaneously by projecting
the received signal vector onto the subspace
orthogonal to that spanned by the vectors
$\mathbf{H}_{k,k}(:,1),\ldots,\mathbf{H}_{k,k}(:,m-1),\mathbf{H}_{k,k}(:,m+1),\ldots,\mathbf{H}_{k,k}(:,N_{{\rm t},k})$, and the column vectors of $\mathbf{H}_{k,k_1},\ldots,\mathbf{H}_{k,k_{L_{ k}}}$. 
Let $\mathbf{U}_k(m)$ be the null space of these column vectors;
the dimension of $\mathbf{U}_k(m)$ 
is $N_{\rm r}\times (N_{\rm r}-(N_{{\rm t},k}-1)-\sum_{i=1}^{L_k}N_{{\rm t},k_i})$ with probability 1.\footnote{$N_{{\rm t},k}-1$ comes from
the dimension of the subspace spanned by
$\mathbf{H}_{k,k}(:,1),\ldots,\mathbf{H}_{k,k}(:,m-1),\mathbf{H}_{k,k}(:,m+1),\ldots,\mathbf{H}_{k,k}(:,N_{{\rm t},k})$ and
$\sum_{i=1}^{L_k}N_{{\rm t},k_i}$ from the dimension
of $\mathbf{H}_{k,k_1},\ldots,\mathbf{H}_{k,k_{L_{ k}}}$.}
By definition of $L_k$, $N_{\rm r}-(N_{{\rm t},k}-1)-\sum_{i=1}^{L_k}N_{{\rm t},k_i}\geq 1$. 

We are interested in maximizing the desired signal power by choosing $\mathbf{v}_{ k}(m)$ in $\mathbf{U}_k(m)$. More precisely, we design $\mathbf{v}_{ k}(m)$ which maximizes $|\mathbf{v}_{ k}^*(m){\mathbf{H}}_{k,k}(:,m)|^2$. If the columns of $\mathbf{U}_k(m)$ are orthonormal bases of the null space, then the following filter maximizes $|\mathbf{v}_{ k}^*(m){\mathbf{H}}_{k,k}(:,m)|^2$:
\begin{align}
\mathbf{v}_{ k}(m)=\frac{\mathbf{U}_k(m)\mathbf{U}_k^*(m)\mathbf{H}_{k,k}(:,m)}{\|\mathbf{U}_k(m)\mathbf{U}_k^*(m)\mathbf{H}_{k,k}(:,m)\|_2}\mbox{.}
\end{align}
By applying this filter, $I_{k1}(m)$ and $I_{k2}(m)$ are suppressed and the resulting SINR becomes
\begin{align}
\mbox{SINR}_k^{\rm ZF}(m)=\frac{H_{k,k}(m)d_{k,k}^{-\alpha}}{I_k(m)+\frac{N_{{\rm t},k}\sigma^2}{P}}\mbox{,}
\end{align}
where $I_k(m)=I_{k3}(m)=\sum_{l,X_l\in\{\Phi\setminus X_k,X_{k_1},X_{k_2},\ldots,X_{k_{L_{ k}}}\}}H_{k,l}d_{k,l}^{-\alpha}$ and $H_{k,k}(m)=\|\mathbf{v}_{ k}^*(m)\mathbf{H}_{k,k}\|^2$ is a Chi-squared random variable\footnote{The probability density function of the Chi-square distribution with $2n$ degrees of freedom, $\mathcal{X}_{2n}^2$, is $f_{\mathcal{X}_{2n}^2}(x)=\frac{x^{n-1}e^{-x}}{(n-1)!}$.} with $2(N_{\rm r}-N_{{\rm t},k}-\sum_{l=1}^{L_{ k}}N_{{\rm t},k_l}+1)$ degrees of freedom \cite{tse2005fundamentals} and $H_{k,l}=\|\mathbf{v}_{ k}^*(m)\mathbf{H}_{k,l}\|^2$ is distributed as a Chi-squared with $2N_{{\rm t},l}$ degrees of freedom \cite{tse2005fundamentals}. The sum spectral efficiency per unit area hence becomes
{\begin{align}\label{eq:palm_zf}
C^{\rm ZF}=
%\frac{1}{|\mathcal{A}|}C_{\mathcal{A}}^{\rm ZF}=
\lambda C_{\rm link}^{\rm ZF}=\lambda\mathbb{E}^0\left[\sum_{m=1}^{N_{{\rm t},0}}\log_2(1+\mbox{SINR}_{0}^{\rm ZF}(m))\right]\mbox{.}
\end{align}}

\subsubsection{ZF-SIC detection} We now consider ZF-SIC,
which is a well-known non-linear detection method for open-loop MIMO systems.
The key idea of ZF-SIC decoding is to recover the data streams successively
and to subtract the recovered streams for obtaining the remaining data streams.
This provides a power gain as well as an interference cancellation gain.
For decoding the data streams of the $k$-th link, the receiver first decodes the signals from interferers using LCSIR. After subtracting off these signals, the $m$-th data of the $k$-th link can be obtained iteratively by decoding and subtracting from the 1st to the $m-1$-th data streams and by then applying the $m$-th decorrelator which suppresses the signal from the $m+1$-th to the $N_{{\rm t},k}$-th streams of the $k$-th link. In other words, the corresponding projection is onto the subspace orthogonal to $\mathbf{H}_{k,k}(:,m+1),\ldots,\mathbf{H}_{k,k}(:,N_{{\rm t},k})$ (say $\tilde{\mathbf{U}}_k(m)$), as opposed to being to the subspace orthogonal to $\mathbf{H}_{k,k}(:,1),\ldots,\mathbf{H}_{k,k}(:,m-1),\mathbf{H}_{k,k}(:,m+1),\ldots,\mathbf{H}_{k,k}(:,N_{{\rm t},k})$ and the column spaces of $\mathbf{H}_{k,k_1},\ldots,\mathbf{H}_{k,k_{L_{ k}}}$ in the previous subsection. By choosing $\tilde{\mathbf{v}}_k(m)$ in $\tilde{\mathbf{U}}_k(m)$ to maximize the signal power, the resulting SINR becomes 
\begin{align}
\mbox{SINR}_k^{\rm SIC}(m)=\frac{\tilde{H}_{k,k}(m)d_{k,k}^{-\alpha}}{\tilde{I}_k(m)+\frac{N_{{\rm t},k}\sigma^2}{P}}\mbox{,}
\end{align} 
where $\tilde{I}_k(m)=I_{k3}(m)=\sum_{l,X_l\in\{\Phi\setminus X_k,X_{k_1},X_{k_2},\ldots,X_{k_{L_{ k}}}\}}\tilde{H}_{k,l}d_{k,l}^{-\alpha}$, $\tilde{H}_{k,k}(m)=\|\tilde{\mathbf{v}}_k^*(m){\mathbf{H}}_{k,k}\|^2$ is a Chi-squared random variable with $2(N_{\rm r}-N_{{\rm t},k}+m)$ degrees of freedom and $\tilde{H}_{k,l}=\|\tilde{\mathbf{v}}_k^*(m)\mathbf{H}_{k,l}\|^2$ is distributed as a Chi-squared with $2N_{{\rm t},l}$ degrees of freedom.\footnote{With the SIC structure, the subspace spanned by $\mathbf{H}_{k,k}(:,m+1),\ldots,\mathbf{H}_{k,k}(:,N_{{\rm t},k})$ is suppressed for recovering the $m$-th data stream.} The sum spectral efficiency per unit area achieved by the ZF-SIC is given by
\begin{align}
C^{\rm SIC}=
%\frac{1}{|\mathcal{A}|}C_{\mathcal{A}}^{\rm SIC}=
\lambda C_{\rm link}^{\rm SIC}=\lambda\mathbb{E}^0\left[\sum_{m=1}^{N_{\rm to}}\log_2(1+\mbox{SINR}_0^{\rm SIC}(m))\right]\mbox{.}
\end{align}
Even though neither ZF nor ZF-SIC are optimal in the information theoretic sense, these are quite commonly used and in addition amenable to analysis. With these receiving architectures, the exact expressions of the sum spectral efficiency and the corresponding scaling laws are given in the following sections.

%%%%%%%%%%%%%%%%%%%%%%%%%%%%%%%%%%%%%%%%%%%%%%%%%%%%%%%%%%%%%%%%%%%%%%%%%%%%%%%%%%%%%%%%%%%
\section{Direct CSIR}\label{sec:Direct_CSIR}
In this section, we obtain the exact analytical expressions of the sum spectral efficiency for both ZF and ZF-SIC detection with DCSIR, i.e., $L_{ k}=0$ for all $X_k\in\Phi$. Then, we derive a lower and an upper bounds with closed-forms. We get the announced scaling laws from these closed from expressions. 

In our closed-form expressions, we use the Gamma function which is defined as $\Gamma(x)=\int_0^{\infty}t^{x-1}e^{-t}dt$.

%First, we consider a case $L_{ k}=0$ for all $k\in\mathcal{K}$ which means a receiver $Y_{ k}$ only knows the channel information from its associated transmitter $X_{ k}$. We refer this case as \emph{direct CSIR}.

\subsection{Sum Spectral Efficiency}

%By leveraging Lemma \ref{lem:useful_lemma} in Appendix \ref{appen:useful lemma}, we propose our main result for the sum spectral efficiency. We denote the sum spectral efficiency per unit area with ZF by $C^{\rm ZF}$ and with ZF-SIC by $C^{\rm SIC}$ in direct CSIR case.

\theorem[ZF with DCSIR]\label{theo:Direct_ZF} 
When using ZF detection,
the sum spectral efficiency per unit area of DCSIR is
\begin{align}\label{eq:ergodic spectral efficiency(ZF)_gen}
C^{\rm ZF}=&\sum_{v=1}^{N_{\rm r}}\frac{\alpha\lambda v p_v}{2\ln 2}\int_{1}^{R_{ d}}\int_{0}^{\infty}\frac
{e^{
-\frac{v \sigma^2 r^\alpha}{P}
\left(\lambda\pi  \sum_{k=1}^{N_{\rm r}}p_k\frac{\Gamma(k+\frac{2}{\alpha})\Gamma(1-\frac{2}{\alpha})}{\Gamma(k)u}\right)^{-\frac{\alpha}{2}}-u}}{u}\nonumber  \\
&\left(1-\left(\frac{1}{1+\left(\lambda\pi r^2 \sum_{k=1}^{N_{\rm r}}p_k\frac{\Gamma(k+\frac{2}{\alpha})\Gamma(1-\frac{2}{\alpha})}{\Gamma(k)u}\right)^{-\frac{\alpha}{2}}}\right)^{N_{\rm r}-v+1}\right)  du    
\frac{2r}{R_d^2-1}dr\mbox{.}
\end{align}
\begin{IEEEproof}
See Appendix \ref{appen:proof_direct_ZF}.
\end{IEEEproof}

\theorem[ZF-SIC with DCSIR]\label{theo:Direct_ZFSIC}
When using ZF-SIC detection,
the sum spectral efficiency per unit area of DCSIR is
\begin{align}\label{eq:ergodic spectral efficiency(ZFSIC)_gen}
 C^{\rm SIC}=&\sum_{v=1}^{N_{\rm r}}{\Bigg[}\frac{\alpha\lambda  p_v}{2\ln 2}\int_{1}^{R_{ d}}\int_{0}^{\infty}
\frac{
e^{
-\frac{v \sigma^2 r^\alpha}{P}}
\left(\lambda\pi  \sum_{k=1}^{N_{\rm r}}p_k\frac{\Gamma(k+\frac{2}{\alpha})\Gamma(1-\frac{2}{\alpha})}{\Gamma(k)u}\right)^{-\frac{\alpha}{2}}-u}{u}\nonumber  \\
&\sum_{m=1}^{v}\left(1-\left(\frac{1}{1+\left(\lambda\pi r^2 \sum_{k=1}^{N_{\rm r}}p_k\frac{\Gamma(k+\frac{2}{\alpha})\Gamma(1-\frac{2}{\alpha})}{\Gamma(k)u}\right)^{-\frac{\alpha}{2}}}\right)^{N_{\rm r}-v+m}\right)  du    
\frac{2r}{R_{ d}^2-1}dr{\Bigg]}\mbox{.}
\end{align}
\begin{IEEEproof}
	See Appendix \ref{appen:proof_direct_ZF}.
\end{IEEEproof}

\corollary When all transmitters have $N_{\rm t}$ antennas,
i.e. $p_{N_{\rm t}}=1$,
\eqref{eq:ergodic spectral efficiency(ZF)_gen} simplifies to
\begin{align}\label{eq:ergodic spectral efficiency(ZF)}
 C^{\rm ZF}=&\frac{\alpha\lambda N_{\rm t}}{2\ln 2}\int_{1}^{R_{ d}}\int_{0}^{\infty}\frac{e^{-\frac{N_{\rm t} \sigma^2 r^\alpha}{P}\left(\lambda\pi  \frac{\Gamma(N_{\rm t}+\frac{2}{\alpha})\Gamma(1-\frac{2}{\alpha})}{\Gamma(N_{\rm t})u}\right)^{-\frac{\alpha}{2}}-u}}{u}\nonumber  \\
&\left(1-\left(\frac{1}{1+\left(\lambda\pi r^2 \frac{\Gamma(N_{\rm t}+\frac{2}{\alpha})\Gamma(1-\frac{2}{\alpha})}{\Gamma(N_{\rm t})u}\right)^{-\frac{\alpha}{2}}}\right)^{N_{\rm r}-N_{\rm t}+1}\right)  du    \frac{2r}{R_{ d}^2-1}dr\mbox{,}
\end{align}
and \eqref{eq:ergodic spectral efficiency(ZFSIC)_gen} reduces to
\begin{align}\label{eq:ergodic spectral efficiency(ZFSIC)}
 C^{\rm SIC}=&\frac{\alpha\lambda }{2\ln 2}\int_{1}^{R_{ d}}\int_{0}^{\infty}\frac{e^{-\frac{N_{\rm t} \sigma^2 r^\alpha}{P}\left(\lambda\pi  \frac{\Gamma(N_{\rm t}+\frac{2}{\alpha})\Gamma(1-\frac{2}{\alpha})}{\Gamma(N_{\rm t})u}\right)^{-\frac{\alpha}{2}}-u}}{u}\nonumber  \\
&\sum_{m=1}^{N_{\rm t}}\left(1-\left(\frac{1}{1+\left(\lambda\pi r^2 \frac{\Gamma(N_{\rm t}+\frac{2}{\alpha})\Gamma(1-\frac{2}{\alpha})}{\Gamma(N_{\rm t})u}\right)^{-\frac{\alpha}{2}}}\right)^{N_{\rm r}-N_{\rm t}+m}\right)  du    \frac{2r}{R_{ d}^2-1}dr\mbox{.}
\end{align}

\begin{figure}[t]
	\begin{subfigure}{0.5\linewidth}
		\begin{center}
			\epsfxsize=3in {\epsfbox{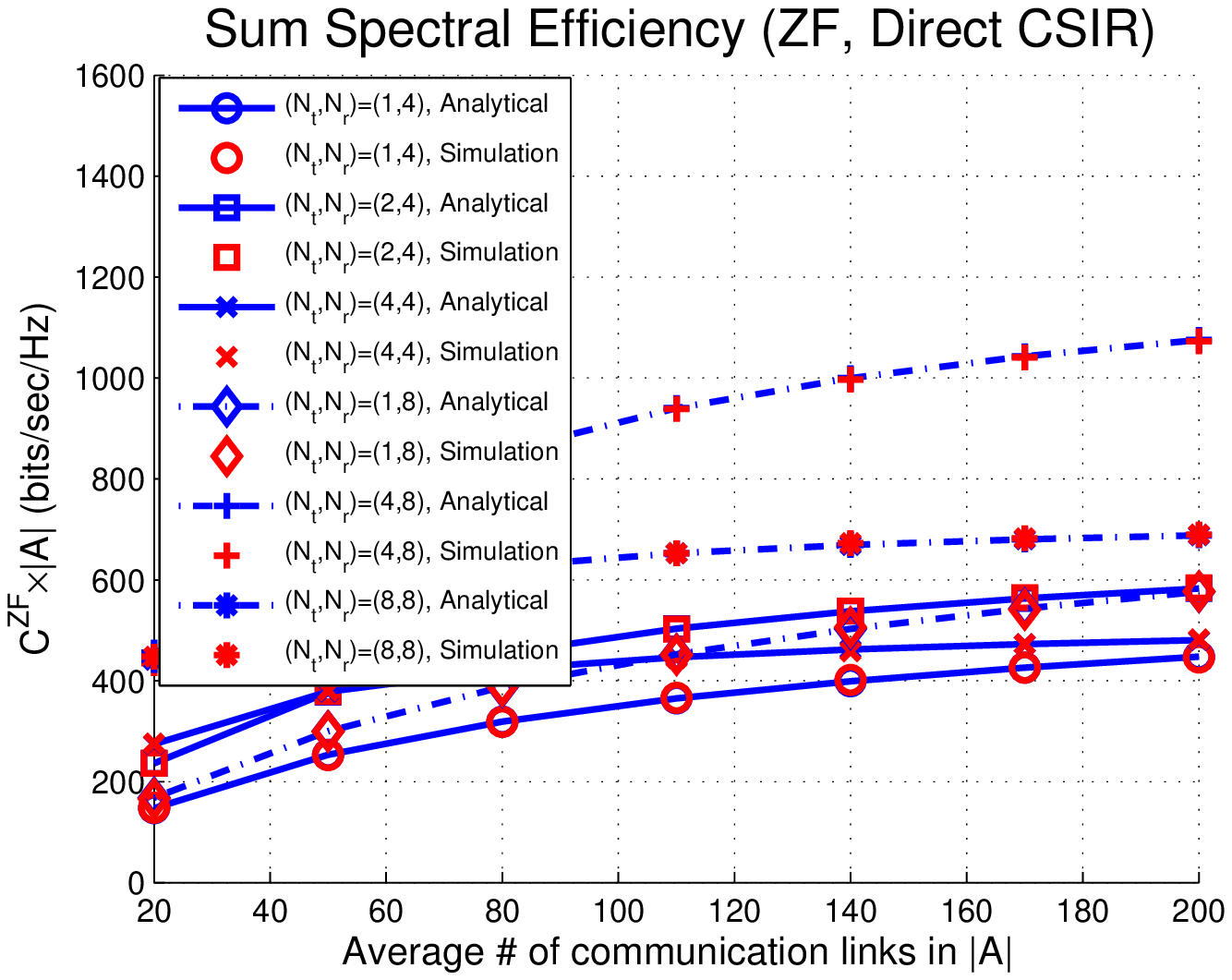}}
			\caption{ZF detection}
		\end{center}
	\end{subfigure}
	\begin{subfigure}{0.5\linewidth}
		\begin{center}
			\epsfxsize=3in {\epsfbox{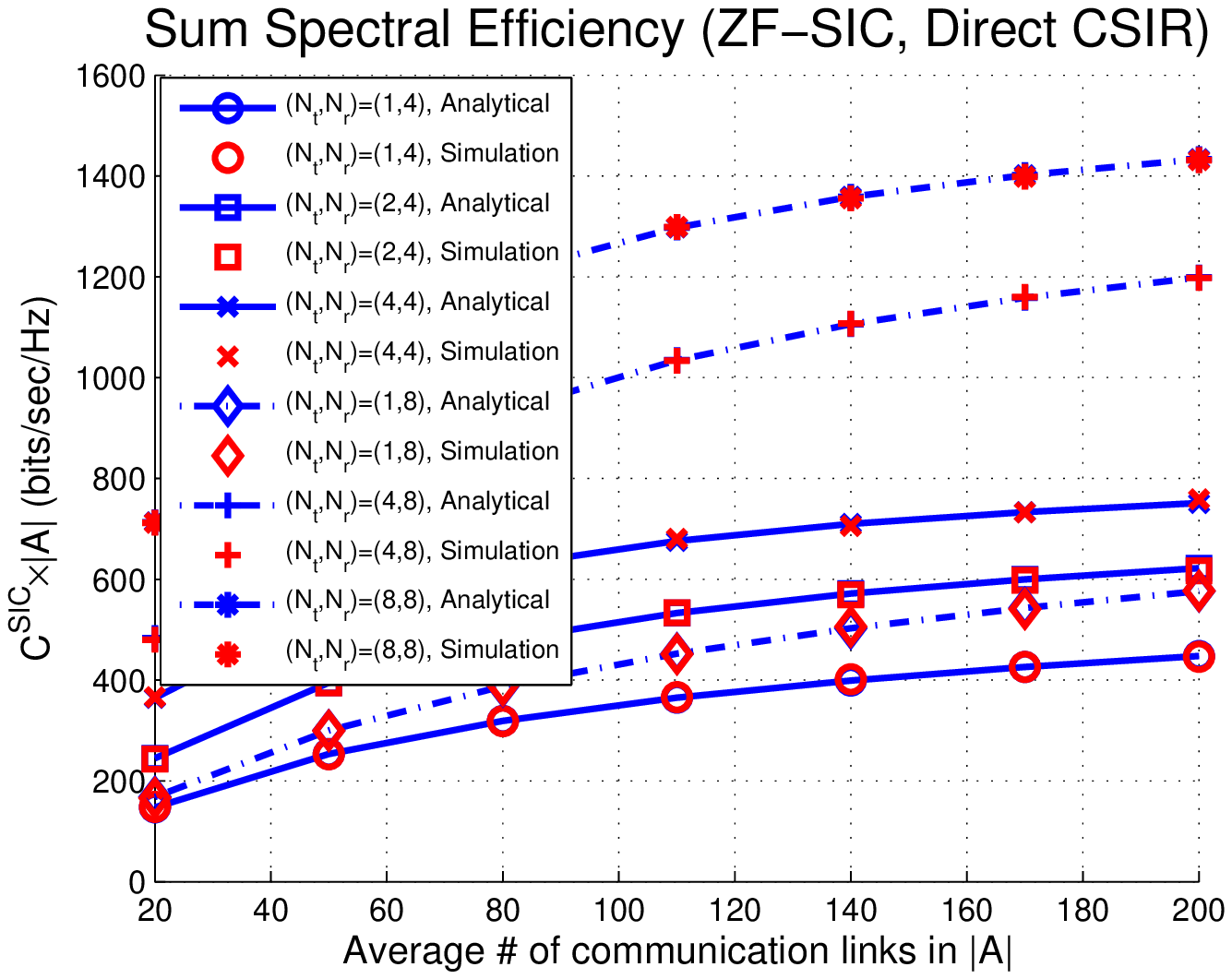}}
			\caption{ZF-SIC detection}
		\end{center}
	\end{subfigure}	
	\caption{\small The sum efficiency with DCSIR when $|\mathcal{A}|=\pi500^2 (m^2)$, $\alpha = 4$, $R_{ d}=50(m)$, $P=-20 (dBm)$, $\sigma^2=-104(dBm)$, $p_{N_{\rm t}}=1$ }
	\label{fig:erg_cap_direct}
\end{figure}

These explicit formulas show how the sum spectral efficiency is determined by the system parameters. Fig.~\ref{fig:erg_cap_direct} plots the sum spectral efficiency of transmitters in region $\mathcal{A}\subset\mathbb{R}^2$ for a path-loss exponent $\alpha=4$, $|\mathcal{A}| = \pi 500^2(m^2)$, $R_{ d} = 50m$, $P=-20dBm$, $p_{N_{\rm t}}=1$ and $\sigma^2=-104dBm$. The gain of the SIC decorrelator can be found by comparing the two figures in Fig.~\ref{fig:erg_cap_direct}.

\remark A drawback of ZF-SIC is error propagation.
In the high SNR regime, however, ZF-SIC detection, which is based
on a higher-dimensional subspace, always provides a higher sum
spectral efficiency than ZF detection, as can be
checked in Fig.~\ref{fig:erg_cap_direct}.

One of the interesting observations is that increasing
the number of streams $N_{\rm t}$ for a given $N_{\rm r}$ and $\lambda$
does not guarantee increasing the sum spectral efficiency. On the one hand,
for a small node density $\lambda$, it is possible to increase
the sum spectral efficiency linearly with the number of spatial multiplexing streams $N_{\rm t}$.
On the other hand, when $\lambda$ is large enough, it is not optimal to send $N_{\rm t}$ data streams,
as the sum spectral efficiency increases sub-linearly with $\lambda$ as shown
in Fig.~\ref{fig:erg_cap_direct}. This implies that, for fixed $N_{\rm t}$ and $N_{\rm r}$,
there exists an optimal density of nodes which maximizes the sum spectral efficiency
per link in such a network. To further obtain insights from the derived expressions,
it is instructive to consider some examples:

\example\label{ex:fixed_dist} When $d_{k,k}=d$ for all $k\in\mathcal{K}$ and $p_{N_{\rm t}}=1$,
Equations \eqref{eq:ergodic spectral efficiency(ZF)} and
\eqref{eq:ergodic spectral efficiency(ZFSIC)} can be simplified as follows 
\begin{align}
C^{\rm ZF} &=\frac{\lambda\alpha N_{\rm t}}{2{\ln}2}\sum_{n=1}^{N_{\rm r}-N_{\rm t}+1}\binom{N_{\rm r}-N_{\rm t}+1}{n}\int_{0}^{\infty}\frac{e^{-u}}{u}\frac{(\frac{\Gamma(N_{\rm t})u}{\lambda \Gamma(N_{\rm t}+\frac{2}{\alpha})\Gamma(1-\frac{2}{\alpha})\pi d^2})^{n\frac{\alpha}{2}}}{(1+(\frac{\Gamma(N_{\rm t})u}{\lambda \Gamma(N_{\rm t}+\frac{2}{\alpha})\Gamma(1-\frac{2}{\alpha})\pi d^2})^{\frac{\alpha}{2}})^{N_{\rm r}-N_{\rm t}+1}}{du}\\
C^{\rm SIC} &=\frac{\lambda\alpha }{2{\ln}2}\sum_{m=1}^{N_{\rm t}}\sum_{n=1}^{N_{\rm r}-N_{\rm t}+m}\binom{N_{\rm r}-N_{\rm t}+m}{n}\int_{0}^{\infty}\frac{e^{-u}}{u}\frac{(\frac{\Gamma(N_{\rm t})u}{\lambda \Gamma(N_{\rm t}+\frac{2}{\alpha})\Gamma(1-\frac{2}{\alpha})\pi d^2})^{n\frac{\alpha}{2}}}{(1+(\frac{\Gamma(N_{\rm t})u}{\lambda \Gamma(N_{\rm t}+\frac{2}{\alpha})\Gamma(1-\frac{2}{\alpha})\pi d^2})^{\frac{\alpha}{2}})^{N_{\rm r}-N_{\rm t}+m}}{du}\mbox{,}
\end{align}
in the interference-limited case ($\sigma^2=0$). This simplified single integral form provides a better intuition on the impact of network design parameters on sum spectral efficiency. For example, increasing $N_{\rm r}$ always provides higher performance, and optimizing $N_{\rm t}$ for fixed $N_{\rm r}$ is an important and interesting question.

\example\label{ex:special_case} Following the Example \ref{ex:fixed_dist}, we further assume that $N_{\rm t} = N_{\rm r}$, $\alpha=4$. In this case, the sum spectral efficiency per unit area with the ZF-receiver is
\begin{align}
C^{\rm ZF}&=\frac{2\lambda N_{\rm t}}{\ln 2}\left\{\sin\left(\frac{\pi\lambda d^2\Gamma(N_{\rm t}+\frac{1}{2})\Gamma(\frac{1}{2})}{\Gamma(N_{\rm t})} \right)  \left(  \frac{\pi}{2}-Si\left(\frac{\pi\lambda d^2\Gamma(N_{\rm t}+\frac{1}{2})\Gamma(\frac{1}{2})}{\Gamma(N_{\rm t})} \right)  \right)\right. \nonumber\\ &\left.-\cos\left(\frac{\pi\lambda d^2\Gamma(N_{\rm t}+\frac{1}{2})\Gamma(\frac{1}{2})}{\Gamma(N_{\rm t})} \right) Ci\left(\frac{\pi\lambda d^2\Gamma(N_{\rm t}+\frac{1}{2})\Gamma(\frac{1}{2})}{\Gamma(N_{\rm t})} \right) \right\}\mbox{,}
\end{align}
where $Si(z)=\int_{0}^{z}\frac{\sin(t)}{t}dt$ and $Ci(z) = -\int_{z}^{\infty}\frac{\cos(t)}{t}dt$ are the sine integral and cosine integral functions.

In Example \ref{ex:special_case}, if we assume $d=\sqrt{\frac{\Gamma(N_{\rm t})}{2\lambda\Gamma(N_{\rm t}+\frac{1}{2})\Gamma(\frac{1}{2})}}$, which means that the distance of communication links is of order of $\lambda^{-\frac{1}{2}}$, the sum spectral efficiency per unit area becomes
\begin{align}
C^{\rm ZF}=\frac{2\lambda N_{\rm t}}{\ln 2}\left(\frac{\pi}{2}-Si\left(\frac{\pi}{2}\right)\right)\simeq0.5772\lambda N_{\rm t}\mbox{.}
\end{align}
So, if the assumptions in Example \ref{ex:special_case} and the above relation of $d$ and $\lambda$ hold, it is possible to guarantee that the sum spectral efficiency per unit area is at least $0.5772N_{\rm t}\lambda$ by choosing $N_{\rm t}$ equal to $N_{\rm r}$.

Throughout this paper, the main scaling is that of the number of transmit and receive antennas with respect to the network density $\lambda$. This example different from the main stream as the link distance depends on the network density $\lambda$. In what follows link distances will not exhibit such a functional depencency.

\subsection{Scaling Law}
In this section, we provide both a lower and an upper bound with a closed-form on
the sum spectral efficiency. This allows us to obtain the announced scaling law.
We focus on the case where $p_{N_{\rm t}}=1$.

\theorem[Direct CSIR, ZF, Scaling Law]\label{theo:ZF_scaling_law}
Assume that $p_{N_{\rm t}}=1$, $N_{\rm t}=c_1\lambda^{\beta_1}$, $N_{\rm r}=c_2\lambda^{\beta_2}$,
for some constants $c_1,c_2>0$, and that $\beta_1\leq\beta_2$.
Then, in the interference limited regime,
%when a ZF-receiver is used, the sum spectral efficiency per unit area scales as
\begin{equation}\label{eq: scaling law_ZF}
\lim_{\lambda \to \infty}{C^{\rm ZF}} = \Theta(\lambda^{\beta_1+1}\log_2(1+\lambda^{\beta_2-\beta_1-\frac{\alpha}{2}}))\mbox{.}
\end{equation}
\begin{IEEEproof}
See Appendix \ref{appen:theo34}.
\end{IEEEproof}

\theorem[Direct CSIR, Scaling Law, ZF-SIC]\label{theo:ZFSIC_scaling_law} Under the same assumptions as in Theorem \ref{theo:ZF_scaling_law}, in the interference limited regime, 
%the sum spectral efficiency per unit area scales as
\begin{equation}\label{eq: scaling law_ZFSIC}
\lim_{\lambda \to \infty}{C^{\rm SIC}} =
\Theta(\lambda^{\beta_1+1}\log_2(1+\lambda^{\beta_2-\beta_1-\frac{\alpha}{2}}))\mbox{.}
\end{equation}
%when ZF-SIC detection is applied.
\begin{IEEEproof}
See Appendix \ref{appen:theo34}.
\end{IEEEproof}	

\remark The first observation is that, in the DCSIR case,
the sum spectral efficiency per unit area are identical for ZF and ZF-SIC in a scaling law sense.
This is because the signal power gain under ZF-SIC is at most $N_{\rm t}$,
i.e., $\mathbb{E}[H_{k,k}(m)]\simeq N_{\rm t}$,
while the fading power of inter-node interference is also
proportional to $\frac{1}{N_{\rm t}}$, i.e.,
($\mathbb{E}\left[\frac{1}{I_k(m)}\right]]\simeq \frac{1}{N_{\rm t}}$).
Consequently, the array gain obtained by ZF-SIC detection is negligible in the scaling law sense.
To obtain a gain from the SIC structure, the signal power gain by ZF-SIC
should be larger than $N_{\rm t}$, and this will actually
be the case for LCISR (see Section \ref{sec:local_CSIR}).

The next corollary, on per link spectral efficiency,
follows immediately from the two theorems stated above.
%We obtain the scaling law of the sum spectral efficiency per unit area as \eqref{eq: scaling law_ZF} and \eqref{eq: scaling law_ZFSIC} when ZF and ZF-SIC detection are used. In the next corollary, we divide the regime of the exponent of transmit and receive antenna with channel exponent and interpret our results in each regime.
\corollary\label{cor:scaling_law_direct} When the receive scheme is
{\rm ZF} or {\rm ZF-SIC}, under DCSIR, the scaling law of the sum
spectral efficiency per link is
\begin{align}
&\Theta(\lambda^{\beta_1}\log(\lambda))~~~&\mbox{for}~\beta_2-\beta_1-\frac{\alpha}{2}>0\mbox{,}\\ 
&\Theta(\lambda^{\beta_1})~~~&\mbox{for}~~\beta_2-\beta_1-\frac{\alpha}{2}=0\mbox{,}\\
&\Theta(\lambda^{\beta_2-\frac{\alpha}{2}})~~~&\mbox{for}~~\beta_2-\beta_1-\frac{\alpha}{2}<0\mbox{.}
\end{align}

%We use the relation of $\log_2(1+x^{\alpha})\simeq \alpha\log_2(x)$ for $\alpha>0$ when $x$ goes to infinity, and $\ln(1+x)\simeq x$ for small $x$.
Here are important observations following from this corollary.
\begin{itemize}
\item
Whenever $\beta_2-\beta_1-\frac{\alpha}{2}\geq 0$, 
the spectral efficiency per link is determined by $N_{\rm t}$ alone.
So, in this regime, \emph{spatial multiplexing}, namely increasing the number
of data streams, is beneficial; 
to the best of our knowledge, this result is new.
\item
Whenever $\beta_2-\beta_1-\frac{\alpha}{2}<0$, 
the sum spectral efficiency
per unit area goes to $0$ exponentially fast with $\lambda$ when $\beta_2<\frac{\alpha}{2}$,
and increases like $\lambda^{\beta_2-\frac{\alpha}{2}}$ 
when $\beta_2>\frac{\alpha}{2}$.
For given $\beta_2$ and $\alpha$ with $\beta_2-\frac{\alpha}{2}>0$,
the best value for $\beta_1$ is $\beta_1^*=\beta_2-\frac{\alpha}{2}$,
and the corresponding scaling law is 
$\Theta(\lambda^{\beta_2-\frac{\alpha}{2}})$.
\item
We can expect a linear gain when $\beta_2=\frac{\alpha}{2}$ as 
this is the critical region between the super-linear
and sub-linear regions.
\item
For fixed $N_{\rm t}$ and $N_{\rm r}$,
(i.e. $\beta_1,\beta_2=0$), the scaling law is
$\Theta(\lambda^{-\frac{\alpha}{2}})$. 
\end{itemize}
\example\label{ex:optimal_density_direct} 
Assume that $p_{N_{\rm t}}=1$.
For fixed values of $N_{\rm t}$, $N_{\rm r}$ and $\alpha$,
what is the optimal node density in our model? We answer this 
question in a heuristic way by
maximizing the lower bounds obtained above.
For the ZF case, the density maximizing the lower bound
of the sum spectral efficiency per unit area in \eqref{eq:lb_zf_direct} is% 
\footnote{Here, we ignore $\epsilon$.}
\begin{align}
\lambda^{*}_{\rm ZF}=\arg\max_{\lambda}\frac{2\lambda N_{\rm t}}{\alpha}\log_2\left(1+\left(\frac{2\Gamma(N_{\rm t})}{\Gamma(N_{\rm t}+\frac{2}{\alpha})\Gamma(1-\frac{2}{\alpha})}\right)^{\frac{\alpha}{2}}\frac{N_{\rm r}-N_{\rm t}}{(\lambda\pi (R_{ d}^2+1))^{\frac{\alpha}{2}}}\right)\mbox{.}
\end{align}
For large $x$, $\log_2(1+x)\simeq\log_2(x)$, so in the high SIR regime, the optimal link density is
\begin{align}
\lambda^{*}_{\rm ZF}=\frac{\Gamma(N_{\rm t})}{2^{\ln 2-1}\Gamma(N_{\rm t}+\frac{2}{\alpha})\Gamma(1-\frac{2}{\alpha})}\frac{(N_{\rm r}-N_{\rm t})^{\frac{2}{\alpha}}}{\pi (R_{ d}^2+1)}\mbox{.}
\end{align}
Hence, the optimal probability in the Aloha protocol
for a given $\lambda$, $N_{\rm t}$, $N_{\rm r}$, $\alpha$ is
\begin{align}
p^{*}_{\rm ZF}=\min(1,\frac{\lambda^{*}_{\rm ZF}}{\lambda})\mbox{.}
\end{align}
For the ZF-SIC case, by using the lower bound
in \eqref{eq:LB_zf_sic_direct} and the relation
$\log_2(1+x)\simeq\log_2(x)$ for large $x$, we get that
the optimal $\lambda$ given $N_{\rm t}$, $N_{\rm r}$,
and $\lambda$ in high SIR regime is
\begin{align}
\lambda^{*}_{ \rm SIC}&=\arg\max_{\lambda}\frac{2\lambda}{\alpha}\sum_{m=1}^{N_{\rm t}}\log_2\left(1+\left(\frac{2\Gamma(N_{\rm t})}{\Gamma(N_{\rm t}+\frac{2}{\alpha})\Gamma(1-\frac{2}{\alpha})}\right)^{\frac{\alpha}{2}}\frac{N_{\rm r}-N_{\rm t}+m-1}{(\lambda\pi (R_{ d}^2+1))^{\frac{\alpha}{2}}}\right)\nonumber\\
&\simeq \arg\max_{\lambda}\frac{2\lambda}{\alpha}\log_2\left(\left(\frac{2\Gamma(N_{\rm t})}{\Gamma(N_{\rm t}+\frac{2}{\alpha})\Gamma(1-\frac{2}{\alpha})\pi (R_{ d}^2+1)}\right)^{\frac{\alpha N_{\rm t}}{2}}\prod_{m=1}^{N_{\rm t}}(N_{\rm r}-N_{\rm t}+m-1)\lambda^{-\frac{\alpha N_{\rm t}}{2}}\right) \nonumber\\
&=\frac{\Gamma(N_{\rm t})}{2^{\ln 2-1}\Gamma(N_{\rm t}+\frac{2}{\alpha})\Gamma(1-\frac{2}{\alpha})\pi (R_{ d}^2+1)}{\left(\prod_{m=1}^{N_{\rm t}}(N_{\rm r}-N_{\rm t}+m-1)\right)^{\frac{2}{N_{\rm t}\alpha}}}\mbox{,}
\end{align} 
and the optimal Aloha probability is
\begin{align}
p^{*}_{\rm SIC}=\min(1,\frac{\lambda^{*}_{\rm SIC}}{\lambda})\mbox{.}
\end{align}

\example Assume $p_{N_{\rm t}}=1$.
For fixed $N_{\rm r}$, $\lambda$, and $\alpha$, What is 
the optimal value for $N_{\rm t}$?
This can be obtained by using the formulas in Theorem \ref{theo:Direct_ZF} and \ref{theo:Direct_ZFSIC}.
A simple way consists in maximizing the lower bounds
as in Example \ref{ex:optimal_density_direct}. By using the Gamma function relation
\begin{align}
\left(\frac{\Gamma(N_{\rm t})}{\Gamma(N_{\rm t}+\frac{2}{\alpha})\Gamma(1-\frac{2}{\alpha})}\right)^{\frac{\alpha}{2}}\geq \frac{1}{N_{\rm t}}\mbox{,}
\end{align}
Equation \eqref{eq:lb_zf_direct}, which is the lower bound of sum spectral efficiency per unit area when ZF-receiver is applied, becomes 
\begin{align}\label{eq:new_lb_zf_direct}
\frac{2\lambda N_{\rm t}}{\alpha}\log_2\left(1+b\frac{N_{\rm r}-N_{\rm t}}{N_{\rm t}}\right)\mbox{,}
\end{align}
when we define
\begin{align}
b\triangleq \left(\frac{2}{\Gamma(1-\frac{2}{\alpha})}\right)^{\frac{\alpha}{2}}\frac{1}{(\lambda\pi R_d^2)^{\frac{\alpha}{2}}}\mbox{.}
\end{align}
In the high SIR regime, the optimal $N_{\rm t}$ for maximizing \eqref{eq:new_lb_zf_direct} is
\begin{align}
N_{\rm t,ZF}^{*}=\frac{bN_{\rm r}}{e}\mbox{.}
\end{align}
In the same manner, we can obtain that the value of $N_{\rm t}$ maximizing \eqref{eq:LB_zf_sic_direct} when ZF-SIC is applied is
\begin{align}
N_{\rm t,SIC}^{*}=N_{\rm t,ZF}^{*}=\frac{bN_{\rm r}}{e}\mbox{.}
\end{align}
%%%%%%%%%%%%%%%%%%%%%%%%%%%%%%%%%%%%%%%%%%%%%%%%%%%%%%%%%%%%%%%%%%%%%%%%%%%%%%%%%%%%%
\section{Local CSIR}\label{sec:local_CSIR}
As already explained, LCSIR denotes the situation
where $L_{ k}>0$, i.e.\, receiver $k$ knows
the $L_{ k}$-nearest interferer CSIs
in addition to the CSI of its own channel.
Through this section, we assume all transmitters are
equipped with $N_{\rm t}$ antennas
(i.e., $p_{N_{\rm t}}=1$) and $L_{ k}=L$ for all $X_k\in\Phi$,
consequently $1\leq L\leq\lfloor\frac{N_{\rm r}}{N_{\rm t}}\rfloor-1$.

\subsection{Sum Spectral Efficiency}
In the LCSIR case,
we denote the sum spectral efficiency per unit area
by $C_{L}^{\rm ZF}$ under ZF and by $C_{L}^{\rm SIC}$ under ZF-SIC.
\theorem\label{theo:Local_ZF} In the LCSIR case,
under ZF detection,
the achievable sum spectral efficiency per unit area with $L$
dominant interferer CSI is
\begin{equation}\label{eq:Local_ZF_sum_spec}
C_{L}^{\rm ZF}=\frac{\lambda N_{\rm t}}{\ln 2}\int_{1}^{R_{ d}}\int_{0}^{\infty}\frac{1}{se^{\frac{N_{\rm t}\sigma^2s}{P}}}\left(1-\frac{1}{(1+sx^{-\alpha})^{N_{\rm r}-(L+1)N_{\rm t}+1}}\right)\mathcal{L}_{\tilde{I}_k}(L;s)ds\frac{2x}{R_{ d}^2}dx,
\end{equation}
where
\begin{equation}
\mathcal{L}_{\tilde{I}_k}(L;s)=\int_{0}^{\infty}\exp\left(-\pi\lambda\int_{u=r^2}^{\infty}1-\left(\frac{1}{1+su^{-\frac{\alpha}{2}}}\right)^{N_{\rm t}}du\right)\frac{2(\lambda\pi r^2)^L}{r\Gamma(L)}e^{-\lambda\pi r^2}dr\mbox{.}
\end{equation}
\begin{IEEEproof}
	See Appendix \ref{appen:proof_local_ZF}.
	\end{IEEEproof}
	
	\theorem\label{theo:Local_ZFSIC} In the local CSIR case, the achievable sum spectral efficiency per unit area with $L$ dominant interferer channel information using ZF-SIC detection is
	\begin{equation}
	C_{L}^{\rm SIC}=\sum_{m=1}^{N_{\rm t}}\frac{\lambda}{\ln 2}\int_{1}^{R_{ d}}\int_{0}^{\infty}\frac{1}{se^{\frac{N_{\rm t}\sigma^2s}{P}}}\left(1-\frac{1}{(1+sx^{-\alpha})^{N_{\rm r}-N_{\rm t}+m}}\right)\mathcal{L}_{\tilde{I}_k}(L;s)ds\frac{2x}{R_{ d}^2}dx,
	\end{equation}
	where
	\begin{equation}
	\mathcal{L}_{\tilde{I}_k}(L;s)=\int_{r=0}^{\infty}\exp\left(-\pi\lambda\int_{u=r^2}^{\infty}1-\left(\frac{1}{1+su^{-\frac{\alpha}{2}}}\right)^{N_{\rm t}}du\right)\frac{2(\lambda\pi r^2)^L}{r\Gamma(L)}e^{-\lambda\pi r^2}dr\mbox{.}
	\end{equation}
	\begin{IEEEproof}
		See Appendix \ref{appen:proof_local_ZF}.
	\end{IEEEproof}

	\begin{figure}[t]
		\begin{subfigure}{0.5\linewidth}
			\begin{center}
				\epsfxsize=3in {\epsfbox{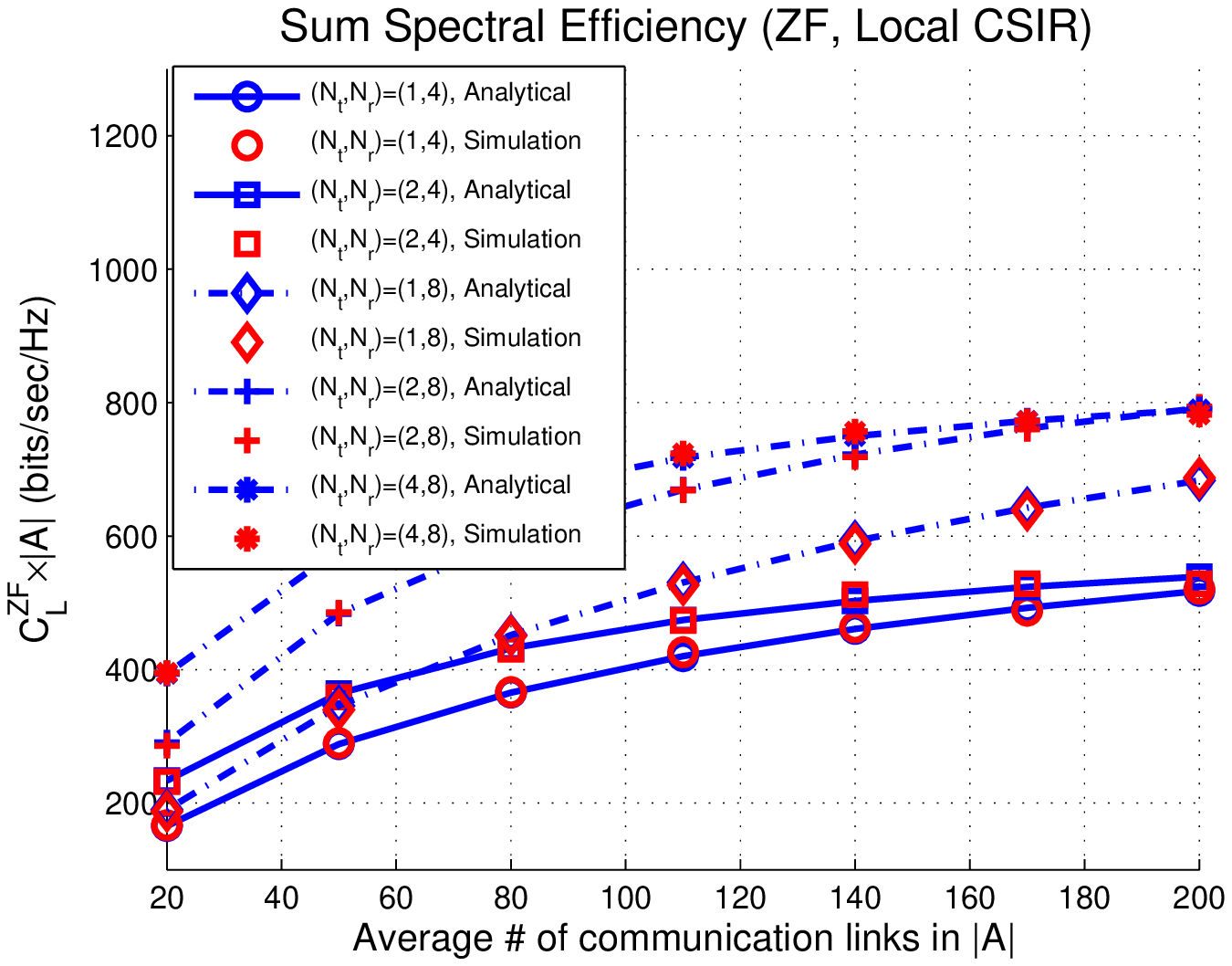}}
				\caption{ZF detection}
			\end{center}
		\end{subfigure}
		\begin{subfigure}{0.5\linewidth}
			\begin{center}
				\epsfxsize=3in {\epsfbox{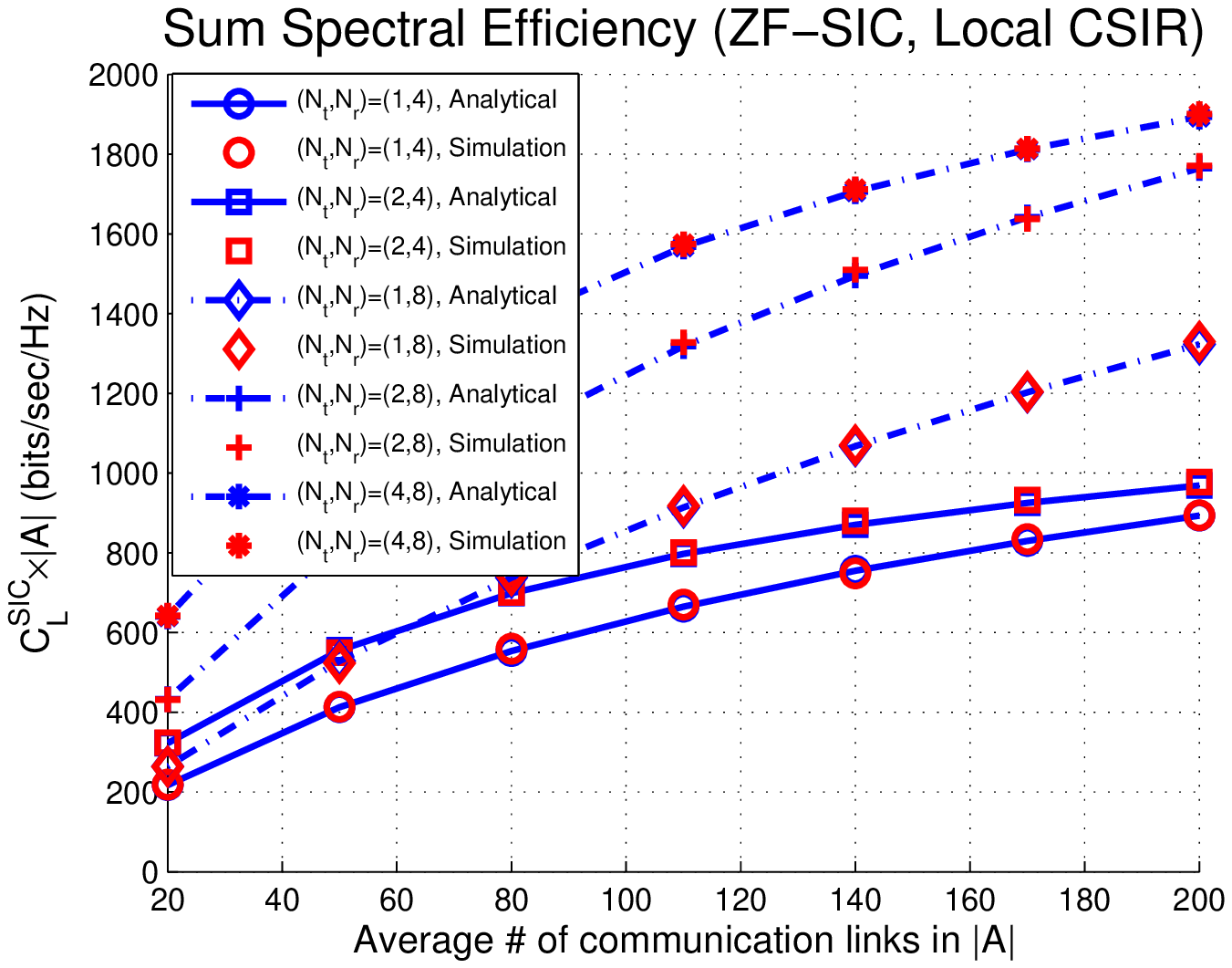}}
				\caption{ZF-SIC detection}
			\end{center}
		\end{subfigure}	
		\caption{\small The sum spectral efficiency with LCSIR when $|\mathcal{A}|=\pi500^2 (m^2)$, $\alpha = 4$, $R_{ d}=50(m)$, $P=-20 (dBm)$, $\sigma^2=-104(dBm)$ with $L=\lfloor\frac{N_{\rm r}}{N_{\rm t}}\rfloor-1$.}
		\label{fig:erg_cap_local}
	\end{figure}
	
Here as in the DCSIR case,
the sum spectral efficiency increases with the network density,
for both ZF and ZF-SIC.
This can be checked in Fig.~\ref{fig:erg_cap_local} where
we see that the sum capacity increases sub-linearly with the
average number of links.
	
\remark For the ZF case, the fading power of the desired signal
is lower for LCSIR than for DCSIR because the remaining degrees
of freedom at the receiver are used to suppress the inter-node
interference from the $L$-dominant interferers. Therefore, leveraging
all channel information is not always beneficial.
This can be checked in the first figures of Fig.~\ref{fig:erg_cap_direct}
and Fig.~\ref{fig:erg_cap_local}.
For the ZF-SIC case, however, utilizing all information is always beneficial,
since the fading power of the $m$-th data stream of the $k$-th link is
$\chi^2_{2(N_{\rm r}-N_{\rm t}+m)}$, rather than $\chi^2_{2(N_{\rm r}-(L+1)N_{\rm t}+1)}$ in ZF.
This observation can be checked on the second figures of
Fig.~\ref{fig:erg_cap_direct} and Fig.~\ref{fig:erg_cap_local}.
%So in the next subsection, the scaling law of ZF detection is obtained with fixed $L$ and the maximum possible $L$, while that of ZF-SIC detection is only considered with maximum possible $L$.

\subsection{Scaling Law}
%As in Section \ref{sec:Direct_CSIR}, we also derive a lower bound of the sum spectral efficiency per unit area which allows ones to prove the scaling law.
In this section, we assume that $L=\lfloor \frac{N_{\rm r}}{N_{\rm t}}\rfloor -1$
which is the maximum possible number for nulling the interference from other nodes.

\theorem[Local CSIR, Scaling Law, ZF]\label{theo:local_sl_zf} Assume that $p_{N_{\rm t}}=1$, and $N_{\rm t}=c_1\lambda^{\beta_1}$, $N_{\rm r}=c_2\lambda^{\beta_2}$ with with some constants $c_1,c_2>0$ and $\beta_1\leq\beta_2$. Then, under ZF detection, the sum spectral efficiency per unit area scales as
\begin{align}\label{eq:Local_scaling_ZF_second}
\lim_{\lambda\rightarrow\infty}{C_{L}^{\rm ZF}}=\Omega(\lambda^{\beta_1+1}\log_2(1+\lambda^{(\beta_2-\beta_1-1)\frac{\alpha}{2}-\beta_2}))\mbox{,}
\end{align}
when $L=\lfloor\frac{N_{\rm r}}{N_{\rm t}}\rfloor-1$.
\begin{IEEEproof}
	See Appendix \ref{appen:theo78}.
\end{IEEEproof}

\remark When $\alpha = 4$, under ZF,
the scaling law of spectral efficiency per link
is $\Theta(\lambda^{\beta_1}\log_2(1+\lambda^{\beta_2-\beta_1-1}))$
for DCSIR, whereas it is $\Omega(\lambda^{\beta_1}\log_2(1+\lambda^{\beta_2-2\beta_1-2}))$ for LCSIR
In this case, we can observe that knowing channel state from other nodes is not useful in the sense of scaling laws.
This is because the receiver wastes the spatial degrees of freedom to cancel the nearest 
inter-node interference. We conclude that, when ZF detection is employed, 
in the scaling law sense,
treating the nearest inter-node interference as noise is a better strategy than canceling it.

\theorem[Local CSIR, Scaling Law, ZF-SIC]\label{theo:local_sl_sic} The assumptions for the number of antenna configurations are the same as in Theorem \ref{theo:local_sl_zf}. When $L=\lfloor\frac{N_{\rm r}}{N_{\rm t}}\rfloor-1$, the sum spectral efficiency per unit area with ZF-SIC detection scales as
\begin{align}\label{eq:Local_scaling_ZFSIC}
\lim_{\lambda\rightarrow\infty}{C_{L}^{\rm SIC}}=\Omega(\lambda^{\beta_1+1}\log_2(1+\lambda^{(\beta_2-\beta_1-1)\frac{\alpha}{2}}))\mbox{.}
\end{align}
\begin{IEEEproof}
	See Appendix \ref{appen:theo78}.
\end{IEEEproof}

The main difference between \eqref{eq:Local_scaling_ZF_second} and \eqref{eq:Local_scaling_ZFSIC} is the degrees of freedom of signal power by the successive cancellation architecture. 

\corollary 
Under {\rm ZF-SIC} and LCSIR,
the scaling law of the ergodic spectral efficiency per link is
\begin{align}
&\Omega(\lambda^{\beta_1}\log(\lambda))~~~&\mbox{for}~\beta_2-\beta_1-1>0\\ 
&\Omega(\lambda^{\beta_1})~~~&\mbox{for}~~\beta_2-\beta_1-1=0\\
&\Omega(\lambda^{\beta_1+(\beta_2-\beta_1-1)\frac{\alpha}{2}})~~~&\mbox{for}~~\beta_2-\beta_1-1<0\mbox{.}
\end{align}

The conclusions are similar to those of Corollary \ref{cor:scaling_law_direct}.
In particular, for given $\beta_2$ and $\alpha$, the best $\beta_1$
in the scaling law sense is hence $\beta_1^*=\beta_2-1$,
and the corresponding scaling law is $\Omega(\lambda^{\beta_2-1})$.
Since we assume $\alpha>2$,
by comparing with the scaling law in Corollary
\ref{cor:scaling_law_direct},
LCSIR can achieve higher performance than
DCSIR case in the ergodic spectral efficiency per link scaling law sense.

\example When $N_{\rm t}$, $N_{\rm r}$, $\lambda$, and $L$ are given,
the density maximizing the lower bounds in \eqref{eq:lb_zf_local}
for ZF and \eqref{eq:lb_sic_local} for ZF-SIC under LCSIR
can be obtained as follows. As in Example \ref{ex:optimal_density_direct},
in the high SIR regime, the optimal densities for ZF and ZF-SIC are
\begin{align}
\lambda^{*}_{\rm ZF,L}&=\left(\frac{N_{\rm r}-(L+1)N_{\rm t}}{\frac{2(1-R_d^{2-\alpha})}{(\alpha-2)(R_d^2-1)}(2\pi)^{\frac{\alpha}{2}}N_{\rm t}}(L-\frac{\alpha}{2})^{\frac{\alpha}{2}-1}\right)^{\frac{2}{\alpha}}\frac{1}{2^{\ln 2}}\\
\lambda^{*}_{\rm SIC,L}&=\left(\frac{(L-\frac{\alpha}{2})^{\frac{\alpha}{2}-1}}{\frac{2(1-R_d^{2-\alpha})}{(\alpha-2)(R_d^2-1)}(2\pi)^{\frac{\alpha}{2}}N_{\rm t}}\right)^{\frac{2}{\alpha}}\left(\prod_{m=1}^{N_{\rm t}}(N_{\rm r}-N_{\rm t}+m-1)\right)^{\frac{2}{N_{\rm t}\alpha}}\frac{1}{2^{\ln 2}} \mbox{,}
\end{align}
and the optimal Aloha probabilities are
\begin{align}
p^{*}_{\rm ZF,L}&=\min\left(1,\frac{\lambda^{*}_{\rm ZF,L}}{\lambda}\right)\\
p^{*}_{\rm SIC,L}&=\min\left(1,\frac{\lambda^{*}_{\rm SIC,L}}{\lambda}\right)\mbox{.}
\end{align}

%%%%%%%%%%%%%%%%%%%%%%%%%%%%%%%%%%%%%%%%%%%%%%%%%%%%%%%%%%%%%%%%%%%%%%%%%%%%%%%%%%%%%%%%%%%
\section{Conclusions}\label{sec:conclu}
We considered a random wireless network with multiple transmit and receive antennas and examined the benefits of using MIMO techniques for obtaining multiplexing gains from the ergodic spectral efficiency point-of-view. Assuming two different types of CSI at receivers, we gave exact analytical expressions and scaling laws for the ergodic spectral efficiency. The main finding is that the ergodic spectral efficiency increases linearly with both the density of nodes and the number of transmit streams, provided that the number of antennas scales in a particular polynomial function with the density. When local CSI with ZF-SIC detection is employed, the lower bound of the scaling law increases
linearly with the density of nodes, the path-loss exponent and the number of transmit antennas provided the ratio between transmit and receive antennas scales in a linear way with the density.  

There are many interesting directions left as future work. One possible direction is to consider antenna correlation effects in both transmit and receive antennas, and to analyze how the correlation effects change the scaling laws. 
Assuming a MIMO random network with finite feedback, it would also be interesting to investigate the benefits of a closed-loop MIMO transmission technique over the open-loop transmission method examined here.
Another direction is to assume a MIMO heterogeneous network and to investigate the optimum number of data streams as a function of the density of nodes.

%%%%%%%%%%%%%%%%%%%%%%%%%%%%%%%%%%%%%%%%%%%%%%%%%%%%%%%%%%%%%%%%%%%%%%%%%%%%%%%%%%%%%%%%%%%

\appendices 

\section{A Lemma for Capacity Analysis}\label{appen:useful lemma}

The following lemma presented in \cite{hamdi2010useful} will be useful below.

\lemma\label{lem:useful_lemma} Let $x_1,\ldots,x_N,y_1,\ldots,y_M$ be arbitrary non-negative random variables. Then
\begin{equation}
\mathbb{E}\left[\ln\left(1+\frac{\sum_{n=1}^Nx_n}{\sum_{m=1}^My_m+1}\right)\right]=\int_{0}^{\infty}\frac{\mathcal{M}_y(z)-\mathcal{M}_{x,y}(z)}{z}\exp(-z)dz,
\end{equation}
where $\mathcal{M}_y(z)=\mathbb{E}\left[e^{-z\sum_{m=1}^My_m}\right]$ and $\mathbb{M}_{x,y}(z)=\mathbb{E}\left[e^{-z(\sum_{n=1}^Nx_n+\sum_{m=1}^My_m)}\right]$.
\begin{IEEEproof}
	See \cite{hamdi2010useful}.
\end{IEEEproof}

The following lemma, proved in
\cite[Appendix B]{lee2016spectral}, will
also be used:
\lemma\label{lem:scale_ineq} Let $X>0$ and $Y>0$ be independent non-negative
random variables such that $\mathbb{E}[X]<\infty$, $\mathbb{E}[Y]<\infty$,
and $\mathbb{E}[\frac{1}{Y}]<\infty$. Then,
\begin{align}
\log_2\left(1+\frac{\exp(\mathbb{E}[\ln(X)])}{\mathbb{E}[Y]}\right)\leq \mathbb{E}_{X,Y}\left[\log_2\left(1+\frac{X}{Y}\right)\right]\leq\log_2\left(1+\mathbb{E}[X]\mathbb{E}\left[\frac{1}{Y}\right]\right)\mbox{.}
\end{align}

\section{Proof of Theorem \ref{theo:Direct_ZF} and \ref{theo:Direct_ZFSIC}}\label{appen:proof_direct_ZF}
Let $X$ and $Y$ be two independent non-negative random variables with $a\in\mathbb{R}^+$, Lemma \ref{lem:useful_lemma} becomes 
\begin{equation}\label{eq:modified_useful_lemma}
\mathbb{E}\left[\ln\left(1+\frac{X}{Y+a}\right)\right]=\int_{0}^{\infty}\frac{e^{-az}}{z}(1-\mathbb{E}[e^{-zX}])\mathbb{E}[e^{-zY}]dz\mbox{.}
\end{equation}
We first prove Theorem \ref{theo:Direct_ZF}. Given $d_{0,0}=d$ for the typical link and $N_{{\rm t},k}=t$, applying \eqref{eq:modified_useful_lemma}, the ergodic spectral efficiency for the $m$-th data stream of the typical link is
\begin{align}\label{eq:pluggedinlemma}
&\mathbb{E}\left[\log_2\left( 1+\frac{H_{0,0}(m)}{d_{0,0}^{\alpha}I_0(m)+\frac{d_{0,0}^\alpha t\sigma^2}{P}}   \right)|d_{0,0}=d, N_{\rm{t}0}=t\right]\nonumber\\=&\frac{1}{\ln 2}\int_{0}^{\infty}\frac{e^{-\frac{d^\alpha t\sigma^2}{P}z}}{z}(1-\mathbb{E}[e^{-zH_{0,0}(m)}])\mathbb{E}[e^{-zd^{\alpha}I_0(m)}]dz\mbox{.}
\end{align}
Let us define $I_0(m)=\bar{I}_{01}(m)+\bar{I}_{02}(m)+\ldots+\bar{I}_{0N_{\rm r}}(m)$, where $\bar{I}_{0k}(m)$ is the interference from nodes which have $k$-transmit antennas. Then, the Laplace transform of the interference $I_0(m)$ is
	\begin{align}
	\mathcal{L}_{I_0(m)}&=\mathbb{E}[e^{-sI_0(m)}]=\mathbb{E}[e^{-s\sum_{k=1}^{N_{\rm r}}\bar{I}_{0k}(m)}]=\prod_{k=1}^{N_{\rm r}}\mathbb{E}[e^{-s\bar{I}_{0k}(m)}]=\prod_{k=1}^{N_{\rm r}}\mathcal{L}_{\bar{I}_{0k}(m)}(s)\mbox{.}
	\end{align}
	
	The Laplace transform of $\bar{I}_{0i}(m)$ is
	\begin{align}
	\mathcal{L}_{\bar{I}_{0k}(m)}(s)&\stackrel{(a)}{=}\exp\left(-\int_{\mathbb{R}^2}\mathbb{E}_{p}[1-e^{-s\frac{p}{r^{\alpha}}}]\lambda p_k dxdy\right)\nonumber\\
	&\stackrel{(b)}{=}\exp\left(-\lambda p_k \int_{0}^{2\pi}\int_{0}^{\infty}\mathbb{E}_p[1-e^{-s\frac{p}{r^{\alpha}}}]rdrd\theta \right)\nonumber\\
	&\stackrel{(c)}{=}\exp\left(-2\pi\lambda p_k \mathbb{E}_p[\int_{0}^{\infty}(1-e^{-s\frac{p}{r^{\alpha}}})rdr]\right)\nonumber\\
	&\stackrel{(d)}{=}\exp\left(-\pi\lambda p_k \mathbb{E}_p[(sp)^{\frac{2}{\alpha}}\int_{0}^{\infty}(1-e^{-u})\frac{-2}{\alpha}\frac{1}{u^{1+\frac{2}{\alpha}}}du]\right)\nonumber\\
	&\stackrel{(e)}{=}\exp\left(-\pi\lambda p_k \mathbb{E}_p[(sp)^{\frac{2}{\alpha}}\int_{0}^{\infty}e^{-u}{u^{-\frac{2}{\alpha}}}du]\right)\nonumber\\
	&\stackrel{(f)}{=}\exp\left(-\pi\lambda p_k \Gamma(1-\frac{2}{\alpha})\mathbb{E}_p[(sp)^{\frac{2}{\alpha}}]\right)\nonumber\\
	&\stackrel{(g)}{=}
\exp\left(-\pi\lambda p_k s^{\frac{2}{\alpha}} \Gamma(1-\frac{2}{\alpha})\frac{\Gamma(k+\frac{2}{\alpha})}{\Gamma(k)}\right)\mbox{.}\nonumber
	\end{align}
	(a) comes from the thinning, the displacement theorem, and the independent marking of PPP \cite{baccelli2009stochastic}; $p$ is the inter-node interference power when $\mathbf{v}_0(m)$ is applied. (b) is obtained by changing from Cartesian coordinates to polar coordinates; (c) is by Fubini's theorem. (d) follows from the change of variable $u=\frac{sp}{r^{\alpha}}$; (e) comes from the integration by part; (f) is by the definition of the Gamma function and (g) comes from the fact that $p$ is a chi-squared random variable with $2k$ degrees of freedom.

So, the Laplace transform of the interference $I_0(m)$ at $zd^{\alpha}$ is
\begin{equation}\label{eq:intfr_2Nt}
\mathbb{E}[e^{-zd^{\alpha}I_0(m)}]=\prod_{k=1}^{N_{\rm r}}\exp\left(-\pi\lambda p_k d^2 z^{\frac{2}{\alpha}}\frac{\Gamma(k+\frac{2}{\alpha})}{\Gamma(k)}\Gamma(1-\frac{2}{\alpha})\right)\mbox{,}
\end{equation}
which comes from the independent thinning and the superposition of PPP with probability generating functional (PGFL) of PPP \cite{baccelli2009stochastic}.
%since $I_k(m)$ experience a Chi-squared with $2N_{\rm t}$ degrees of freedom \cite{baccelli2009stochastic}. 
By plugging \eqref{eq:intfr_2Nt} into \eqref{eq:pluggedinlemma}, we obtain
\begin{align}
\scriptsize&\scriptsize\frac{1}{\ln 2}\int_{0}^{\infty}\frac{e^{-\frac{d^\alpha t\sigma^2}{P}z}}{z}(1-\mathbb{E}[e^{-zH_{0,0}(m)}])\exp\left(-\pi\lambda d^2 z^{\frac{2}{\alpha}}\sum_{k=1}^{N_{\rm r}}p_k\frac{\Gamma(k+\frac{2}{\alpha})}{\Gamma(k)}\Gamma(1-\frac{2}{\alpha})\right)dz\nonumber\\
\scriptsize\stackrel{(a)}{=}&\scriptsize\frac{\alpha}{2\ln 2}\int_{0}^{\infty}\frac{e^{-\frac{d^{\alpha}t \sigma^2}{P}\left(\lambda\pi d^2 \sum_{k=1}^{N_{\rm r}}p_k\frac{\Gamma(k+\frac{2}{\alpha})\Gamma(1-\frac{2}{\alpha})}{\Gamma(k)u}\right)^{-\frac{\alpha}{2}}}}{u}  \nonumber\\
&\times\left(1-\mathbb{E}\left[e^{-\left(\lambda\pi d^2 \sum_{k=1}^{N_{\rm r}}p_k\frac{\Gamma(k+\frac{2}{\alpha})\Gamma(1-\frac{2}{\alpha})}{\Gamma(k)u}\right)^{-\frac{\alpha}{2}}H_{k,k}(m)}\right]\right)  e^{-u}du\nonumber\\\label{eq:proof_ZF_ZFSIC_diff}
\scriptsize\stackrel{(b)}{=}&\scriptsize\frac{\alpha}{2\ln 2}\int_{0}^{\infty}\frac{e^{-\frac{d^{\alpha}t \sigma^2}{P}\left(\lambda\pi d^2 \sum_{k=1}^{N_{\rm r}}p_k\frac{\Gamma(k+\frac{2}{\alpha})\Gamma(1-\frac{2}{\alpha})}{\Gamma(k)u}\right)^{-\frac{\alpha}{2}}-u}}{u} \nonumber\\
&\times\left(1-\left(\frac{1}{1+\left(\lambda\pi d^2 \sum_{k=1}^{N_{\rm r}}p_k\frac{\Gamma(k+\frac{2}{\alpha})\Gamma(1-\frac{2}{\alpha})}{\Gamma(k)u}\right)^{-\frac{\alpha}{2}}}\right)^{N_{\rm r}-t+1}\right)  du\mbox{,}
\end{align}
where (a) comes from a variable change, and (b) follows from deconditioning $H_{k,k}(m)$ which is a Chi-squared random variable with
$2(N_{\rm r}-t+1)$ degrees of freedom.
Since $Y_{ k}$ is uniformly distributed in the ring centered at $X_{k}$, we obtain \eqref{eq:ergodic spectral efficiency(ZF)} by considering all data streams and deconditioning w.r.t. the number of transmit antennas of the typical link.

For the ZF-SIC detection method, the main difference in the proof is that $\tilde{H}_{0,0}(m)$ is distributed as a Chi-squared with 
$2(N_{\rm r}-t+m)$ degrees of freedom,
and \eqref{eq:proof_ZF_ZFSIC_diff} is changed to 
\begin{align}
&\frac{\alpha}{2\ln 2}\int_{0}^{\infty}\frac{e^{-\frac{d^{\alpha}t \sigma^2}{P}\left(\lambda\pi d^2 \sum_{k=1}^{N_{\rm r}}p_k\frac{\Gamma(k+\frac{2}{\alpha})\Gamma(1-\frac{2}{\alpha})}{\Gamma(k)u}\right)^{-\frac{\alpha}{2}}-u}}{u} \nonumber\\&\left(1-\left(\frac{1}{1+\left(\lambda\pi d^2 \sum_{k=1}^{N_{\rm r}}p_k\frac{\Gamma(k+\frac{2}{\alpha})\Gamma(1-\frac{2}{\alpha})}{\Gamma(k)u}\right)^{-\frac{\alpha}{2}}}\right)^{N_{\rm r}-t+m}\right)  du\mbox{,}
\end{align}
and we obtain \eqref{eq:ergodic spectral efficiency(ZFSIC)} similarly.

%\section{Proof of Theorem \ref{theo:Direct_ZFSIC}}\label{appen:proof_direct_ZFSIC}

\section{Proof of Theorem \ref{theo:ZF_scaling_law} and \ref{theo:ZFSIC_scaling_law}}\label{appen:theo34}

\begin{IEEEproof}
		We start to derive the lower and upper bounds of \eqref{eq:ergodic spectral efficiency(ZF)}. By applying Lemma \ref{lem:scale_ineq}, the 
		sum spectral efficiency over the network is lower bounded as follows:
		\begin{align}
		\lambda\mathbb{E}^0\left[\sum_{m=1}^{N_{\rm t}}\log_2(1+\mbox{SINR}_0^{\rm ZF}(m))\right]&=\lambda\sum_{m=1}^{N_{\rm t}}\mathbb{E}_{H_{0,0}(m),d_{0,0},I_0(m)}\left[\log_2\left(1+\frac{H_{0,0}(m)d_{0,0}^{-\alpha}}{I_0(m)}\right)\right]\nonumber\\&\geq\lambda\sum_{m=1}^{N_{\rm t}}\mathbb{E}_{d_{0,0},I_0(m)}\left[\log_2\left(1+\frac{e^{\mathbb{E}[\ln(H_{0,0}(m))]}}{d_{0,0}^{\alpha}I_0(m)}\right)\right]\mbox{.}
		\end{align}
		Since $H_{0,0}(m)$ is a Chi-square random variable with $2(N_{\rm r}-N_{\rm t}+1)$ degrees of freedom, 
		\begin{align}
		\mathbb{E}[\ln(H_{0,0}(m))]=\psi(N_{\rm r}-N_{\rm t}+1)\mbox{,}
		\end{align}
		where 
		\begin{align}
		\psi(n)=-\gamma+\sum_{j=1}^{n-1}\frac{1}{j}\mbox{,}
		\end{align}
		with $\gamma \simeq 0.577$, Euler's constant. By \cite[Theorem 3.1]{laforgia2013some},
		\begin{align}\label{eq:ineq_exp_psi}
		e^{\psi(x)}> x-1\mbox{,}
		\end{align}
		and we obtain
		\begin{align}
		e^{\mathbb{E}[\ln(H_{0,0}(m))]}> N_{\rm r}-N_{\rm t}+\epsilon\mbox{,}
		\end{align}
		where $\epsilon$ is some positive number\footnote{With a numerical approach, the gap of $e^{\psi(x)}$ and $x-1$ is lower bounded by 0.4. For obtaining lower bound of the sum spectral efficiency (and scaling law of it), we just put $\epsilon$ to prevent the lower bound becoming $0$.}.
		Thus, the lower bound of the sum spectral efficiency per unit area is	
		\begin{align}\label{eq:lb_zf_direct}
		&\lambda\sum_{m=1}^{N_{\rm t}}\mathbb{E}_{d_{0,0},I_0(m)}\left[\log_2\left(1+\frac{N_{\rm r}-N_{\rm t}+\epsilon}{d_{0,0}^{\alpha}I_0(m)}\right)\right]\nonumber\\&\stackrel{(a)}{=}\frac{\lambda}{\ln 2}\sum_{m=1}^{N_{\rm t}}\int_{0}^{\infty}\frac{1}{z}(1-e^{-z(N_{\rm r}-N_{\rm t}+\epsilon)})\mathbb{E}_{d_{0,0},I_0(m)}[e^{-zd_{0,0}^{\alpha}I_0(m)}]dz\nonumber\\
		&\stackrel{(b)}{=}\frac{\lambda}{\ln 2}\sum_{m=1}^{N_{\rm t}}\int_{0}^{\infty}\frac{1}{z}(1-e^{-z(N_{\rm r}-N_{\rm t}+\epsilon)})\mathbb{E}_{d_{0,0}}\left[\exp\left(-\lambda\pi d_{0,0}^2z^{\frac{2}{\alpha}}\frac{\Gamma(N_{\rm t}+\frac{2}{\alpha})\Gamma(1-\frac{2}{\alpha})}{\Gamma(N_{\rm t})}\right)\right]dz\nonumber\\
		&\stackrel{(c)}{\geq}\frac{\lambda}{\ln 2}\sum_{m=1}^{N_{\rm t}}\int_{0}^{\infty}\frac{1}{z}(1-e^{-z(N_{\rm r}-N_{\rm t}+\epsilon)})\exp\left(-\lambda\pi \mathbb{E}[d_{0,0}^2]z^{\frac{2}{\alpha}}\frac{\Gamma(N_{\rm t}+\frac{2}{\alpha})\Gamma(1-\frac{2}{\alpha})}{\Gamma(N_{\rm t})}\right)dz\nonumber\\
		&\stackrel{(d)}{=}\frac{\lambda\alpha}{2\ln 2}\sum_{m=1}^{N_{\rm t}}\int_{0}^{\infty}\frac{1}{u}e^{-u}\left(1-e^{-\left(\frac{2\Gamma(N_{\rm t})}{\Gamma(N_{\rm t}+\frac{2}{\alpha})\Gamma(1-\frac{2}{\alpha})}\right)^{\frac{\alpha}{2}}\frac{N_{\rm r}-N_{\rm t}+\epsilon}{(\lambda\pi (R_{ d}^2+1))^{\frac{\alpha}{2}}}u^{\frac{\alpha}{2}}}\right)du\nonumber\\
		&\stackrel{(e)}{\geq} \frac{\lambda}{\ln 2}\sum_{m=1}^{N_{\rm t}}\int_{0}^{\infty}\frac{1}{u}e^{-u^{\frac{\alpha}{2}}}\left(1-e^{-\left(\frac{2\Gamma(N_{\rm t})}{\Gamma(N_{\rm t}+\frac{2}{\alpha})\Gamma(1-\frac{2}{\alpha})}\right)^{\frac{\alpha}{2}}\frac{N_{\rm r}-N_{\rm t}+\epsilon}{(\lambda\pi (R_{ d}^2+1))^{\frac{\alpha}{2}}}u^{\frac{\alpha}{2}}}\right)du\nonumber\\
		&\stackrel{(f)}{=}\frac{2\lambda N_{\rm t}}{\alpha}\log_2\left(1+\left(\frac{2\Gamma(N_{\rm t})}{\Gamma(N_{\rm t}+\frac{2}{\alpha})\Gamma(1-\frac{2}{\alpha})}\right)^{\frac{\alpha}{2}}\frac{N_{\rm r}-N_{\rm t}+\epsilon}{(\lambda\pi (R_{ d}^2+1))^{\frac{\alpha}{2}}}\right)\nonumber\\
		&\stackrel{(g)}{\geq}\frac{2\lambda N_{\rm t}}{\alpha}\log_2\left(1+\frac{1}{N_{\rm t}}\left(\frac{2}{\Gamma(1-\frac{2}{\alpha})}\right)^{\frac{\alpha}{2}}\frac{N_{\rm r}-N_{\rm t}+\epsilon}{(\lambda\pi (R_{ d}^2+1))^{\frac{\alpha}{2}}}\right)\mbox{,}
		\end{align}
		where (a) follows from Lemma \ref{lem:useful_lemma}, (b) comes from the expression for the interference of the Laplace functional of PPP, (c) follows from Lemma \ref{lem:scale_ineq}, (d) comes from a variable change and the fact that $\mathbb{E}[d_{k,k}^2]=\frac{R_{ d}^2+1}{2}$, (e) comes from the fact that $e^{-u}\geq \frac{2}{\alpha}e^{-u^{\frac{\alpha}{2}}}$ when $u\geq 0$ and $\alpha>2$, (f) is obtained by $\int_{0}^{\infty}\frac{1}{u}e^{-u^{\frac{\alpha}{2}}}(1-e^{-b\times u^{\frac{\alpha}{2}}})du=\frac{2}{\alpha}\log(1+b)$, and (g) comes from 	\begin{align}
		\frac{\Gamma(N_{\rm t})}{\Gamma(N_{\rm t}+\frac{2}{\alpha})}\geq N_{\rm t}^{-\frac{2}{\alpha}}\mbox{.}
		\end{align}
		Using the assumption that $N_{\rm t}=c_1\lambda^{\beta_1}$ and $N_{\rm r}=c_2\lambda^{\beta_2}$, we obtain	
		\begin{align}\label{eq:zf_direct_omega_exp}
		\lim_{\lambda\rightarrow\infty}	{C^{\rm ZF}} = \Omega(\lambda^{\beta_1+1}\log_2(1+\lambda^{\beta_2-\beta_1-\frac{\alpha}{2}}))\mbox{,}
		\end{align}
		as $\lambda$ goes to infinity.
		
		Next, we derive an upper bound when the receiver applies ZF. In the interference limited regime, 
		\begin{align}\label{eq:ub_zf_direct}
		\lambda\mathbb{E}&\left[\sum_{m=1}^{N_{\rm t}}\log_2(1+\mbox{SINR}_0^{\rm ZF}(m))\right]=\lambda\sum_{m=1}^{N_{\rm t}}\mathbb{E}_{H_{0,0}(m),d_{0,0},I_0(m)}\left[\log_2\left(1+\frac{H_{0,0}(m)d_{0,0}^{-\alpha}}{I_0(m)}\right)\right]\nonumber\\
		&\stackrel{(a)}{\leq}\lambda N_{\rm t}\log_2\left(1+\mathbb{E}[d_{0,0}^{-\alpha}]\mathbb{E}[H_{0,0}(m)]\mathbb{E}\left[\frac{1}{I_0(m)}\right]\right)\nonumber\\
		&\stackrel{(b)}{=}\lambda N_{\rm t}\log_2\left(1+\frac{2(1-R_d^{2-\alpha})}{(\alpha-2)(R_d^2-1)}\frac{1}{R_{ d}^{\alpha}}(N_{\rm r}-N_{\rm t}+1)\frac{\Gamma(1+\frac{2}{\alpha})\Gamma(N_{\rm t})^{\frac{\alpha}{2}}}{(\lambda\pi\Gamma(N_{\rm t}+\frac{2}{\alpha})\Gamma(1-\frac{2}{\alpha}))^{\frac{\alpha}{2}}}\right)\mbox{,}
		\end{align}where (a) comes from Lemma \ref{lem:scale_ineq}, and (b) follows from $\mathbb{E}[d_{0,0}^{-\alpha}]=\frac{2(1-R_d^{2-\alpha})}{(\alpha-2)(R_d^2-1)}$, $\mathbb{E}[H_{0,0}(m)]=N_{\rm r}-N_{\rm t}+1$, and the relation of $\mathbb{E}\left[\frac{1}{X}\right]=\mathbb{E}\left[\int_{0}^{\infty}e^{-sX}ds\right]$ for any positive random variable $X$. The negative moment of $I_0(m)$ is
		\begin{align}
		\mathbb{E}\left[\frac{1}{I_0(m)}\right]=\int_{0}^{\infty}\mathbb{E}[e^{-sI_0(m)}]=\int_{0}^{\infty}e^{-\lambda\pi\frac{\Gamma(N_{\rm t}+\frac{2}{\alpha})\Gamma(1-\frac{2}{\alpha})}{\Gamma(N_{\rm t})}s^{\frac{2}{\alpha}}}ds=\frac{\Gamma(1+\frac{2}{\alpha})\Gamma(N_{\rm t})^{\frac{\alpha}{2}}}{(\lambda\pi\Gamma(N_{\rm t}+\frac{2}{\alpha})\Gamma(1-\frac{2}{\alpha}))^{\frac{\alpha}{2}}}\mbox{.}
		\end{align}
		Therefore, the upper bound on the sum spectral efficiency per unit area is
		\begin{align}
		{C^{\rm ZF}}&=\lambda N_{\rm t}\log_2\left(1+\frac{2(1-R_d^{2-\alpha})}{(\alpha-2)(R_d^2-1)}(N_{\rm r}-N_{\rm t}+1)\frac{\Gamma(1+\frac{2}{\alpha})\Gamma(N_{\rm t})^{\frac{\alpha}{2}}}{(\lambda\pi\Gamma(N_{\rm t}+\frac{2}{\alpha})\Gamma(1-\frac{2}{\alpha}))^{\frac{\alpha}{2}}}\right)\nonumber\\
		&\leq \lambda N_{\rm t}\log_2\left(1+\frac{2(1-R_d^{2-\alpha})}{(\alpha-2)(R_d^2-1)}(N_{\rm r}-N_{\rm t}+1)\frac{\Gamma(1+\frac{2}{\alpha})}{(\lambda\pi\Gamma(1-\frac{2}{\alpha}))^{\frac{\alpha}{2}}}\left((N_{\rm t}-1)^{-\frac{2}{\alpha}}\right)^{\frac{\alpha}{2}}\right)\mbox{,}
		\end{align}
		where the last inequality comes from 
		\begin{align}\label{eq:ineq_1}
		\frac{\Gamma(x)}{\Gamma(x+\frac{2}{\alpha})}\leq(x-1)^{-\frac{2}{\alpha}}\mbox{.}
		\end{align}
		By letting $\lambda$ tend to infinity, we obtain
		\begin{align}\label{eq:zf_direct_o_exp}
		\lim_{\lambda\rightarrow\infty}{C^{\rm ZF}}=\mathcal{O}(\lambda^{\beta_1+1}\log_2(1+\lambda^{\beta_2-\beta_1-\frac{\alpha}{2}}))\mbox{.}
		\end{align}
		Equations \eqref{eq:zf_direct_omega_exp} and \eqref{eq:zf_direct_o_exp} conclude the proof of Theorem \ref{theo:ZF_scaling_law}.

		The proof of Theorem \ref{theo:ZFSIC_scaling_law} is analogous to that of Theorem \ref{theo:ZF_scaling_law}. The main difference consists in changing $H_{0,0}(m)~\sim\chi_{2(N_{\rm r}-N_{\rm t}+1)}^2$ to $\tilde{H}_{0,0}(m)\sim\chi^2_{2(N_{\rm r}-N_{\rm t}+m)}$. The lower bound of the sum spectral efficiency per unit area becomes
		\begin{align}
		&\lambda\mathbb{E}\left[\sum_{m=1}^{N_{\rm t}}\log_2\left(1+\mbox{SINR}_{0}^{\rm SIC}(m)\right)\right]&\nonumber\\
		&>\frac{2\lambda}{\alpha}\sum_{m=1}^{N_{\rm t}}\log_2\left(1+\left(\frac{2\Gamma(N_{\rm t})}{\Gamma(N_{\rm t}+\frac{2}{\alpha})\Gamma(1-\frac{2}{\alpha})}\right)^{\frac{\alpha}{2}}\frac{{N_{\rm r}-N_{\rm t}+m-1+\epsilon}}{(\lambda\pi (R_{ d}^2+1))^{\frac{\alpha}{2}}}\right)&\nonumber\\
		&>\frac{2\lambda N_{\rm t}}{\alpha}\log_2\left(1+\left(\frac{2}{\pi (R_{ d}^2+1)\Gamma(1-\frac{2}{\alpha})}\right)^{\frac{\alpha}{2}}\frac{{N_{\rm r}-N_{\rm t}+\epsilon}}{N_{\rm t}}\lambda^{-\frac{\alpha}{2}}\right)\mbox{,}&
		\end{align}
		and the upper bound becomes
		\begin{align}\label{eq:LB_zf_sic_direct}
		&\lambda\mathbb{E}\left[\sum_{m=1}^{N_{\rm t}}\log_2\left(1+\mbox{SINR}_{0}^{\rm SIC}(m)\right)\right]\nonumber\\
		&\leq\lambda\sum_{m=1}^{N_{\rm t}}\log_2\left(1+\frac{2(1-R_d^{2-\alpha})}{(\alpha-2)(R_d^2-1)}(N_{\rm r}-N_{\rm t}+m)\frac{\Gamma(1+\frac{2}{\alpha})\Gamma(N_{\rm t})^{\frac{\alpha}{2}}}{(\lambda\pi\Gamma(N_{\rm t}+\frac{2}{\alpha})\Gamma(1-\frac{2}{\alpha}))^{\frac{\alpha}{2}}}\right)\nonumber\\
		&<\lambda{N_{\rm t}}\log_2\left(1+\frac{2(1-R_d^{2-\alpha})}{(\alpha-2)(R_d^2-1)}\frac{\Gamma(1+\frac{2}{\alpha})}{(\pi\Gamma(1-\frac{2}{\alpha}))^{\frac{\alpha}{2}}}N_{\rm r}\left(\frac{\Gamma(N_{\rm t})}{\Gamma(N_{\rm t}+\frac{2}{\alpha})}\right)^{\frac{\alpha}{2}}\lambda^{-\frac{\alpha}{2}}\right)\nonumber\\
		&\leq\lambda{N_{\rm t}}\log_2\left(1+\frac{2(1-R_d^{2-\alpha})}{(\alpha-2)(R_d^2-1)}\frac{\Gamma(1+\frac{2}{\alpha})}{(\pi\Gamma(1-\frac{2}{\alpha}))^{\frac{\alpha}{2}}}\frac{N_{\rm r}}{N_{\rm t}-1}\lambda^{-\frac{\alpha}{2}}\right)\mbox{,}
		\end{align}
		where the last inequality comes from \eqref{eq:ineq_1}. With the foregoing assumptions, the scaling law of the sum spectral per unit area with respect to the density becomes $\Theta(\lambda^{\beta_1+1}\log_2(1+\lambda^{\beta_2-\beta_1-\frac{\alpha}{2}}))$.
\end{IEEEproof}
\section{Proof of Theorem \ref{theo:Local_ZF} and \ref{theo:Local_ZFSIC}}\label{appen:proof_local_ZF}

We use Lemma \ref{lem:useful_lemma} again. We start to derive the ZF-receiver case. Conditioned on $d_{k,k}=d$, the spectral efficiency of the $m$-th data stream of the typical link is
\begin{align}\label{eq:pluggedinlemma_SIC}
\mathbb{E}\left[\log_2\left( 1+\frac{\tilde{H}_{0,0}(m)d_{0,0}^{-\alpha}}{\tilde{I}_0(m)+\frac{ N_{\rm t}\sigma^2}{P}}   \right)|d_{0,0}=d\right]&=\frac{1}{\ln 2}\int_{0}^{\infty}\frac{e^{-\frac{ N_{\rm t}\sigma^2}{P}z}}{z}(1-\mathbb{E}[e^{-zH_{0,0}(m)}d^{-\alpha}])\nonumber\\
&\times\mathbb{E}[e^{-z\sum_{j=L+1}^{\infty}\tilde{H}_{0,0_j}(m)d_{0,0_j}^{-\alpha}}]dz\mbox{,}
\end{align}
by Lemma \ref{lem:useful_lemma}. Since $\tilde{H}_{0,0}(m)$ is Chi-square distributed with $2(N_{\rm r}-(L+1)N_{\rm t}+1)$ distributed,
\begin{equation}
\mathbb{E}\left[e^{-z\tilde{H}_{0,0}(m)d^{-\alpha}}\right]=\frac{1}{(1+zd^{-\alpha})^{N_{\rm r}-(L+1)N_{\rm t}+1}}\mbox{.}
\end{equation}
The Laplace transform of $\tilde{I}_0(m)$ for the given $L$ is
\begin{equation}
\mathcal{L}_{\tilde{I}_0(m)}(L;s)=\mathbb{E}\left[e^{-z\sum_{j=L+1}^{\infty}\tilde{H}_{0,0_j}(m)d_{0,0_j}^{-\alpha}}\right]\mbox{.}
\end{equation}
Under the condition that the $L$-th nearest interferer's
distance is $r$, the Laplace transform is obtained as
\begin{align}
\mathcal{L}_{\tilde{I}_0|d_{0,L}=r}(L;s)&=\mathbb{E}\left[e^{-z\sum_{j=L+1}^{\infty}\tilde{H}_{0,0_j}(m)d_{0,0_j}^{-\alpha}}|\{d_{0,L}=r\}\right]\nonumber\\
&\stackrel{(a)}{=}\mathbb{E}\left[\prod_{d_{0,0_j}\in\Phi\setminus\mathcal{B}(0,r)}\frac{1}{(1+zd_{0,0_j}^{-\alpha})^{N_{\rm t}}}|\{d_{0,L}=r\} \right]\nonumber\\
&\stackrel{(b)}{=}\exp\left(-\pi\lambda\int_{u=r^2}^{\infty}1-\frac{1}{(1+zu^{-\frac{\alpha}{2}})^{N_{\rm t}}}du\right)\mbox{,}
\end{align}
where (a) comes from the fact that $\tilde{H}_{0,0_j}(m)\sim\chi^2_{2N_{\rm t}}$ and (b) follows from PGFL.
The distribution of $r$ is given in \cite{haenggi2005distances} and by unconditioning with respect to it, 
\begin{align}
\mathcal{L}_{\tilde{I}_0}(L;s)=\int_{0}^{\infty}\exp\left(-\pi\lambda\int_{u=r^2}^{\infty}1-\frac{1}{(1+zu^{-\frac{\alpha}{2}})^{N_{\rm t}}}du\right)\frac{2(\lambda\pi r^2)^L}{r\Gamma(L)}e^{-\lambda\pi r^2}dr\mbox{.}
\end{align}
Thus, the sum spectral efficiency conditioned on $d_{k,k}=d$ can be written as
\begin{align}
&\mathbb{E}\left[\log_2\left(1+\frac{\tilde{H}_{0,0}(m)d^{-\alpha}}{\tilde{I}_0(m)+\frac{\sigma^2 N_{\rm t}}{P}}\right)|\{d_{0,0}=d\}\right]\nonumber\\
&=\frac{1}{\ln 2}\int_{0}^{\infty}\frac{e^{-\frac{sN_{\rm t}\sigma^2}{P}}}{s}\left[1-\frac{1}{(1+zd^{-\alpha})^{N_{\rm r}-(L+1)N_{\rm t}+1}}\right]\mathcal{L}_{\tilde{I}_0}(L;s) ds\mbox{.}
\end{align}
We obtain the announced result when using the fact that
$d_{0,0}$ is uniformly distributed in
a ring with radii $(1,R_{ d})$.

The result for ZF-SIC follows by the same arguments,
using the fact that $\tilde{H}_{0,0}(m)$ is Chi-square
random variable with $2(N_{\rm r}-N_{\rm t}+m)$ degrees of freedom.

\section{Proof of Theorem \ref{theo:local_sl_zf} and \ref{theo:local_sl_sic}}\label{appen:theo78}
\begin{IEEEproof}
	We start the proof of Theorem \ref{theo:local_sl_zf}.
	The lower bound of \eqref{eq:Local_ZF_sum_spec} is 
	\begin{align}
	\lambda\mathbb{E}\left[\sum_{m=1}^{N_{\rm t}}\log_2(1+\mbox{SINR}_{0,L}^{\rm ZF}(m))\right]&=\lambda\sum_{m=1}^{N_{\rm t}}\mathbb{E}_{\tilde{H}_{0,0}d_{0,0},\tilde{I}_0(m)}\left[\log_2\left(1+\frac{\tilde{H}_{0,0}(m)d_{0,0}^{-\alpha}}{\tilde{I}_0(m)}\right)\right]\nonumber\\
	&\stackrel{(a)}{\geq} \lambda\sum_{m=1}^{N_{\rm t}}\log_2\left(1+\frac{e^{\mathbb{E}[\ln(\tilde{H}_{0,0}(m))]}}{\mathbb{E}[d_{0,0}^{-\alpha}]\mathbb{E}[\tilde{I}_0(m)]}\right)\nonumber\\
	&\stackrel{(b)}{>} \lambda\sum_{m=1}^{N_{\rm t}}\log_2\left(1+\frac{N_{\rm r}-(L+1)N_{\rm t}+\epsilon}{\frac{2(1-R_d^{2-\alpha})}{(\alpha-2)(R_d^2-1)}\mathbb{E}[\tilde{I}_k(m)]}\right)\mbox{,}
	\end{align}
	where $(a)$ comes from Lemma \ref{lem:scale_ineq}, and $(b)$ comes from the inequality \eqref{eq:ineq_exp_psi}, $\mathbb{E}[d_{0,0}^{-\alpha}]=\frac{2(1-R_d^{2-\alpha})}{(\alpha-2)(R_d^2-1)}$. The expectation of $\tilde{I}_0$ conditioned on $d_{0,0_L}=r$ is 
	\begin{align}
	\mathbb{E}[\tilde{I}_0(m)|d_{0,0_L}=r]=\frac{2\pi\lambda N_{\rm t}}{2-\alpha}r^{2-\alpha}\mbox{.}
	\end{align}
	By unconditioning with respect to $d_{0,0_L}$ whose distribution is given in \cite{haenggi2005distances}, we get
	\begin{align}
	\mathbb{E}[\tilde{I}_0(m)]&=\frac{2\pi\lambda N_{\rm t}}{2-\alpha}\int_{0}^{\infty}r^{2-\alpha} \frac{2(\lambda\pi r^2)^L}{r\Gamma(L)} e^{-\lambda\pi r^2}   dr\\
	&=(2\pi\lambda)^{\frac{\alpha}{2}}N_{\rm t}\frac{\Gamma(1-\frac{\alpha}{2}+L)}{\Gamma(L)}\mbox{.}
	\end{align}
	By leveraging 
	\begin{align}
	\frac{\Gamma(L)}{\Gamma(1-\frac{\alpha}{2}+L)}\geq(L-\frac{\alpha}{2})^{\frac{\alpha}{2}-1}\mbox{,}
	\end{align}
	the lower bound becomes
	\begin{align}\label{eq:lb_zf_local}
	\lambda\mathbb{E}\left[\sum_{m=1}^{N_{\rm t}}\log_2(1+\mbox{SINR}_{0,L}^{\rm ZF}(m))\right]&>\lambda N_{\rm t}\log_2\left(1+\frac{N_{\rm r}-(L+1)N_{\rm t}+\epsilon}{\frac{2(1-R_d^{2-\alpha})}{(\alpha-2)(R_d^2-1)}}\frac{\Gamma(L)}{(2\pi\lambda)^{\frac{\alpha}{2}}N_{\rm t}\Gamma(1-\frac{\alpha}{2}+L)}\right)\nonumber\\
	&\geq \lambda N_{\rm t}\log_2\left(1+\frac{N_{\rm r}-(L+1)N_{\rm t}+\epsilon}{\frac{2(1-R_d^{2-\alpha})}{(\alpha-2)(R_d^2-1)}(2\pi\lambda)^{\frac{\alpha}{2}}N_{\rm t}}(L-\frac{\alpha}{2})^{\frac{\alpha}{2}-1}\right)\mbox{.}	
	\end{align}
	By plugging $N_{\rm t}=c_1\lambda^{\beta_1}$, $N_{\rm r}=c_2\lambda^{\beta_2}$ into \eqref{eq:lb_zf_local}, we obtain the following scaling law:
	%\begin{align}
	%\lim_{\lambda\rightarrow\infty}\frac{C_{\sum,L}^{\rm ZF}}{\lambda}	&=\Omega(\lambda^{\beta_1}\log_2(1+\lambda^{\beta_2-\beta_1-\frac{\alpha}{2}}))\mbox{,}
	%	\end{align}
	%with fixed $L$ and for $L=\lfloor \frac{N_{\rm r}}{N_{\rm t}} \rfloor -1$, we obtain
	\begin{align}
	\lim_{\lambda\rightarrow\infty}{C_{L}^{\rm ZF}}&=\Omega(\lambda^{\beta_1+1}\log_2(1+\lambda^{(\beta_2-\beta_1-1)\frac{\alpha}{2}-\beta_2}))\mbox{,}
	\end{align}
	since $L=\lfloor\frac{N_{\rm r}}{N_{\rm t}}\rfloor-1$.

	The proof of Theorem \ref{theo:local_sl_sic} is almost identical to the proof of Theorem \ref{theo:local_sl_zf}. The main difference is in the distribution of $\tilde{H}_{0,0}(m)$. The lower bound becomes 
	\begin{align}\label{eq:lb_sic_local}
	\lambda\mathbb{E}\left[\sum_{m=1}^{N_{\rm t}}\log_2(1+\mbox{SINR}_{0,L}^{\rm SIC}(m))\right]>\lambda\sum_{m=1}^{N_{\rm t}}\log_2\left(1+\frac{N_{\rm r}-N_{\rm t}+m-1+\epsilon}{\frac{2(1-R_d^{2-\alpha})}{(\alpha-2)(R_d^2-1)}(2\pi\lambda)^{\frac{\alpha}{2}}N_{\rm t}}(L-\frac{\alpha}{2})^{\frac{\alpha}{2}-1}\right)\mbox{.}
	\end{align}
	With the foregoing assumptions, we obtain
	\begin{align}
	\lim_{\lambda\rightarrow\infty}{C_{L}^{\rm SIC}}=\Omega(\lambda^{\beta_1+1}\log_2(1+\lambda^{(\beta_2-\beta_1-1)\frac{\alpha}{2}}))\mbox{.}
	\end{align}
\end{IEEEproof}

\section*{Acknowledgement}{\small 
	This work is supported in part by the National Science Foundation under Grant No. NSF-CCF-1218338 and an award from the Simons Foundation $(\# 197982)$, both to the University of Texas at Austin.}

\bibliographystyle{Bibs/ieeetran}
\bibliography{Bibs/IEEEabrv,Bibs/referenceBibs}

\begin{thebibliography}{10}
\providecommand{\url}[1]{#1}
\csname url@samestyle\endcsname
\providecommand{\newblock}{\relax}
\providecommand{\bibinfo}[2]{#2}
\providecommand{\BIBentrySTDinterwordspacing}{\spaceskip=0pt\relax}
\providecommand{\BIBentryALTinterwordstretchfactor}{4}
\providecommand{\BIBentryALTinterwordspacing}{\spaceskip=\fontdimen2\font plus
\BIBentryALTinterwordstretchfactor\fontdimen3\font minus
  \fontdimen4\font\relax}
\providecommand{\BIBforeignlanguage}[2]{{%
\expandafter\ifx\csname l@#1\endcsname\relax
\typeout{** WARNING: IEEEtran.bst: No hyphenation pattern has been}%
\typeout{** loaded for the language `#1'. Using the pattern for}%
\typeout{** the default language instead.}%
\else
\language=\csname l@#1\endcsname
\fi
#2}}
\providecommand{\BIBdecl}{\relax}
\BIBdecl

\bibitem{baccelli2009stochasticvol2}
F.~Baccelli and B.~Blaszczyszyn, \emph{Stochastic Geometry and Wireless
  Networks: Volume 2: APPLICATIONS}.\hskip 1em plus 0.5em minus 0.4em\relax Now
  Publishers Inc, 2009, vol.~2.

\bibitem{gupta2000capacity}
P.~Gupta and P.~R. Kumar, ``The capacity of wireless networks,'' \emph{IEEE
  Transactions on Information Theory}, vol.~46, no.~2, pp. 388--404, 2000.

\bibitem{toumpis2003capacity}
S.~Toumpis and A.~J. Goldsmith, ``Capacity regions for wireless ad hoc
  networks,'' \emph{IEEE Transactions on Wireless Communications}, vol.~2,
  no.~4, pp. 736--748, 2003.

\bibitem{hartenstein2008tutorial}
H.~Hartenstein and K.~P. Laberteaux, ``A tutorial survey on vehicular ad hoc
  networks,'' \emph{IEEE Communications Magazine}, vol.~46, no.~6, pp.
  164--171, 2008.

\bibitem{doppler2009device}
K.~Doppler, M.~Rinne, C.~Wijting, C.~B. Ribeiro, and K.~Hugl,
  ``Device-to-device communication as an underlay to {LTE}-advanced networks,''
  \emph{IEEE Communications Magazine}, vol.~47, no.~12, pp. 42--49, 2009.

\bibitem{fodor2012design}
G.~Fodor, E.~Dahlman, G.~Mildh, S.~Parkvall, N.~Reider, G.~Mikl{\'o}s, and
  Z.~Tur{\'a}nyi, ``Design aspects of network assisted device-to-device
  communications,'' \emph{IEEE Communications Magazine}, vol.~50, no.~3, pp.
  170--177, 2012.

\bibitem{blum2003mimo}
R.~S. Blum, ``{MIMO} capacity with interference,'' \emph{IEEE Journal on
  Selected Areas in Communications}, vol.~21, no.~5, pp. 793--801, 2003.

\bibitem{chen2006mimo}
B.~Chen and M.~J. Gans, ``{MIMO} communications in ad hoc networks,''
  \emph{IEEE Transactions on Signal Processing}, vol.~54, no.~7, pp.
  2773--2783, 2006.

\bibitem{ozgur2007hierarchical}
A.~{\"O}zg{\"u}r, O.~L{\'e}v{\^e}que, and D.~N. Tse, ``Hierarchical cooperation
  achieves optimal capacity scaling in ad hoc networks,'' \emph{IEEE
  Transactions on Information Theory}, vol.~53, no.~10, pp. 3549--3572, 2007.

\bibitem{tulino2004random}
A.~M. Tulino and S.~Verd{\'u}, \emph{Random matrix theory and wireless
  communications}.\hskip 1em plus 0.5em minus 0.4em\relax Now Publishers Inc,
  2004, vol.~1.

\bibitem{yu2005capacity}
X.~Yu, R.~M. De~Moraes, H.~Sadjadpour, and J.~Garcia-Luna-Aceves, ``Capacity of
  {MIMO} mobile wireless ad hoc networks,'' in \emph{International Conference
  on Wireless Networks, Communications and Mobile Computing, 2005},
  vol.~2.\hskip 1em plus 0.5em minus 0.4em\relax IEEE, 2005, pp. 1053--1058.

\bibitem{franceschetti2007closing}
M.~Franceschetti, O.~Dousse, N.~David, and P.~Thiran, ``Closing the gap in the
  capacity of wireless networks via percolation theory,'' \emph{IEEE
  Transactions on Information Theory}, vol.~53, no.~3, pp. 1009--1018, 2007.

\bibitem{leveque2004information}
O.~L{\'e}v{\^e}que and E.~Telatar, ``Information theoretic upper bounds on the
  capacity of large extended ad-hoc wireless networks,'' in \emph{Proceedings
  of the 2004 IEEE International Symposium on Information Theory}, no.
  LTHI-CONF-2006-010, 2004.

\bibitem{grossglauser2001mobility}
M.~Grossglauser and D.~Tse, ``Mobility increases the capacity of ad-hoc
  wireless networks,'' in \emph{INFOCOM 2001. Twentieth Annual Joint Conference
  of the IEEE Computer and Communications Societies. Proceedings. IEEE},
  vol.~3.\hskip 1em plus 0.5em minus 0.4em\relax IEEE, 2001, pp. 1360--1369.

\bibitem{negi2004capacity}
R.~Negi and A.~Rajeswaran, ``Capacity of power constrained ad-hoc networks,''
  in \emph{INFOCOM 2004. Twenty-third AnnualJoint Conference of the IEEE
  Computer and Communications Societies}, vol.~1.\hskip 1em plus 0.5em minus
  0.4em\relax IEEE, 2004.

\bibitem{franceschetti2009capacity}
M.~Franceschetti, M.~D. Migliore, and P.~Minero, ``The capacity of wireless
  networks: Information-theoretic and physical limits,'' \emph{IEEE
  Transactions on Information Theory}, vol.~55, no.~8, pp. 3413--3424, 2009.

\bibitem{baccelli2009stochastic}
F.~Baccelli and B.~Blaszczyszyn, \emph{Stochastic Geometry and Wireless
  Networks: Volume 1: THEORY}.\hskip 1em plus 0.5em minus 0.4em\relax Now
  Publishers Inc, 2009, vol.~1.

\bibitem{Stoyan}
D.~Stoyan, W.~S. Kendall, and J.~Mecke, \emph{Stochastic Geometry and its
  Applications}, 2nd~ed.\hskip 1em plus 0.5em minus 0.4em\relax Chichester:
  Wiley, 1995.

\bibitem{weber2005transmission}
S.~P. Weber, X.~Yang, J.~G. Andrews, and G.~De~Veciana, ``Transmission capacity
  of wireless ad hoc networks with outage constraints,'' \emph{IEEE
  Transactions on Information Theory}, vol.~51, no.~12, pp. 4091--4102, 2005.

\bibitem{andrews2007ad}
J.~G. Andrews, S.~Weber, and M.~Haenggi, ``Ad hoc networks: to spread or not to
  spread?[ad hoc and sensor networks],'' \emph{IEEE Communications Magazine},
  vol.~45, no.~12, pp. 84--91, 2007.

\bibitem{weber2007transmission}
S.~P. Weber, J.~G. Andrews, X.~Yang, and G.~De~Veciana, ``Transmission capacity
  of wireless ad hoc networks with successive interference cancellation,''
  \emph{IEEE Transactions on Information Theory}, vol.~53, no.~8, pp.
  2799--2814, 2007.

\bibitem{blomer2009transmission}
J.~Blomer and N.~Jindal, ``Transmission capacity of wireless ad hoc networks:
  Successive interference cancellation vs. joint detection,'' in \emph{IEEE
  International Conference on Communications, 2009. ICC'09.}\hskip 1em plus
  0.5em minus 0.4em\relax IEEE, 2009, pp. 1--5.

\bibitem{zhang2014performance}
X.~Zhang and M.~Haenggi, ``The performance of successive interference
  cancellation in random wireless networks,'' \emph{IEEE Transactions on
  Information Theory}, vol.~60, no.~10, pp. 6368--6388, 2014.

\bibitem{hunter2008transmission}
A.~M. Hunter, J.~G. Andrews, and S.~Weber, ``Transmission capacity of ad hoc
  networks with spatial diversity,'' \emph{IEEE Transactions on Wireless
  Communications}, vol.~7, no.~12, pp. 5058--5071, 2008.

\bibitem{jindal2011multi}
N.~Jindal, J.~G. Andrews, and S.~Weber, ``Multi-antenna communication in ad hoc
  networks: Achieving {MIMO} gains with {SIMO} transmission,'' \emph{IEEE
  Transactions on Communications}, vol.~59, no.~2, pp. 529--540, 2011.

\bibitem{akoum2011spatial}
S.~Akoum, M.~Kountouris, M.~Debbah, and R.~W. Heath, ``Spatial interference
  mitigation for multiple input multiple output ad hoc networks: {MISO}
  gains,'' in \emph{IEEE 2011 Conference Record of the Forty Fifth Asilomar
  Conference on Signals, Systems and Computers (ASILOMAR)}.\hskip 1em plus
  0.5em minus 0.4em\relax IEEE, 2011, pp. 708--712.

\bibitem{louie2011open}
R.~H. Louie, M.~R. McKay, and I.~B. Collings, ``Open-loop spatial multiplexing
  and diversity communications in ad hoc networks,'' \emph{IEEE Transactions on
  Information Theory}, vol.~57, no.~1, pp. 317--344, 2011.

\bibitem{huang2012spatial}
K.~Huang, J.~G. Andrews, D.~Guo, R.~W. Heath, and R.~A. Berry, ``Spatial
  interference cancellation for multiantenna mobile ad hoc networks,''
  \emph{IEEE Transactions on Information Theory}, vol.~58, no.~3, pp.
  1660--1676, 2012.

\bibitem{vaze2012transmission}
R.~Vaze and R.~W. Heath, ``Transmission capacity of ad-hoc networks with
  multiple antennas using transmit stream adaptation and interference
  cancellation,'' \emph{IEEE Transactions on Information Theory}, vol.~58,
  no.~2, pp. 780--792, 2012.

\bibitem{kountouris2009transmission}
M.~Kountouris and J.~G. Andrews, ``Transmission capacity scaling of {SDMA} in
  wireless ad hoc networks,'' in \emph{IEEE Information Theory Workshop, 2009.
  ITW 2009.}\hskip 1em plus 0.5em minus 0.4em\relax IEEE, 2009, pp. 534--538.

\bibitem{lee2015spectral}
N.~Lee, D.~Morales-Jimenez, A.~Lozano, and R.~W. Heath, ``Spectral efficiency
  of dynamic coordinated beamforming: A stochastic geometry approach,''
  \emph{IEEE Transactions on Wireless Communications}, vol.~14, no.~1, pp.
  230--241, 2015.

\bibitem{lozano2012yesterday}
A.~Lozano and N.~Jindal, ``Are yesterday-s information-theoretic fading models
  and performance metrics adequate for the analysis of today's wireless
  systems?'' \emph{IEEE Communications Magazine}, vol.~50, no.~11, pp.
  210--217, 2012.

\bibitem{lee2016spectral}
N.~Lee, F.~Baccelli, and R.~W. Heath, ``Spectral efficiency scaling laws in
  dense random wireless networks with multiple receive antennas,'' \emph{IEEE
  Transactions on Information Theory}, vol.~62, no.~3, pp. 1344--1359, 2016.

\bibitem{tse2005fundamentals}
D.~Tse and P.~Viswanath, \emph{Fundamentals of wireless communication}.\hskip
  1em plus 0.5em minus 0.4em\relax Cambridge university press, 2005.

\bibitem{hamdi2010useful}
K.~A. Hamdi, ``A useful lemma for capacity analysis of fading interference
  channels,'' \emph{IEEE Transactions on Communications}, vol.~58, no.~2, pp.
  411--416, 2010.

\bibitem{laforgia2013some}
A.~Laforgia and P.~Natalini, ``On some inequalities for the gamma function,''
  \emph{Advances in Dynamical Systems and Applications}, vol.~8, no.~2, pp.
  261--267, 2013.

\bibitem{haenggi2005distances}
M.~Haenggi, ``On distances in uniformly random networks,'' \emph{IEEE
  Transactions on Information Theory}, vol.~51, no.~10, pp. 3584--3586, 2005.

\end{thebibliography}

\end{document}